\documentclass[fleqn,usenatbib]{mnras}

\usepackage[T1]{fontenc}

\DeclareRobustCommand{\VAN}[3]{#2}
\let\VANthebibliography\thebibliography
\def\thebibliography{\DeclareRobustCommand{\VAN}[3]{##3}\VANthebibliography}

\usepackage{graphicx}	% Including figure files
\usepackage{amsmath}	% Advanced maths commands
\usepackage{amssymb}	% Extra maths symbols
\usepackage{multirow}
\usepackage{pdflscape}
\usepackage{subfigure}
\usepackage[dvipsnames,table,xcdraw]{xcolor}
\newcommand{\cmps}{\ensuremath{\textrm{cm}~\textrm{s}^{-1}}}
\newcommand{\cmpss}{\ensuremath{\textrm{cm}~\textrm{s}^{-2}}}
\newcommand{\mps}{\ensuremath{\textrm{m}~\textrm{s}^{-1}}}
\newcommand{\kmps}{\ensuremath{\textrm{km}~\textrm{s}^{-1}}}
\newcommand{\mystar}{Kepler-37}

\newcommand{\lrhk}{\ensuremath{\log R'_{\rm HK}}}
\newcommand{\bis}{\ensuremath{{\rm BIS}}}
\newcommand{\pchord}{\texttt{PolyChord}}

\newcommand\hl[1]{{#1}}
\newcommand\tocheck[1]{\textcolor{black}{{#1}}}

\newcommand{\bl}{--} % line to indicate blank cell in table
\renewcommand{\th}{\textsuperscript{th}} % ex: I won 4\th place
\usepackage{newtxtext,newtxmath}
\title[A HARPS-N mass for the elusive Kepler-37d]{\hl{A HARPS-N mass for the elusive Kepler-37d: a case study in disentangling stellar activity and planetary signals}}

\author[V.~M.~Rajpaul et al.]{V.~M.~Rajpaul,$^{1,2}$\thanks{E-mail: vr325@cam.ac.uk}
L.~A.~Buchhave,$^{3}$
G.~Lacedelli,$^{4,5}$
K.~Rice,$^{6,7}$ 
A.~Mortier,$^{1,8}$
L.~Malavolta,$^{4,5}$
S.~Aigrain,$^{2}$  \newauthor
L.~Borsato,$^{5}$
A.~W.~Mayo,$^{9}$
D.~Charbonneau,$^{10}$
M.~Damasso,$^{11}$
X.~Dumusque,$^{12}$
A.~Ghedina,$^{13}$\newauthor
D.~W.~Latham,$^{10}$ 
M.~L\'opez-Morales,$^{10}$ 
A.~Magazz\`u,$^{13}$
G.~Micela,$^{14}$
E.~Molinari,$^{15}$
F.~Pepe,$^{12}$ 
G.~Piotto,$^{4,5}$\newauthor
E.~Poretti,$^{13,16}$  
S.~Rowther,$^{17,18}$
A.~Sozzetti,$^{11}$
S.~Udry,$^{12}$ and
C.~A.~Watson$^{19}$
\\
$^{1}$Astrophysics Group, Cavendish Laboratory, JJ Thomson Avenue, Cambridge CB3 0HE , UK\\
$^{2}$Sub-department of Astrophysics, Department of Physics, University of Oxford, Oxford OX1 3RH, UK\\
$^{3}$DTU Space, National Space Institute, Technical University of Denmark, Elektrovej 328, DK-2800 Kgs.\ Lyngby, Denmark\\
$^{4}$Department of Physics and Astronomy, Universit\`a degli Studi di Padova, Vicolo dell'Osservatorio 3, IT-35122 Padova, Italy\\
$^{5}$INAF -- Osservatorio Astronomico di Padova, Vicolo dell'Osservatorio 5, IT-35122 Padova, Italy\\
$^{6}$SUPA, Institute for Astronomy, University of Edinburgh, Blackford Hill, Edinburgh EH9 3HJ, Scotland, UK\\
$^{7}$Centre for Exoplanet Science, University of Edinburgh, Peter Guthrie Tait Road, Edinburgh EH9 3FD, UK\\
$^{8}$Kavli Institute for Cosmology, University of Cambridge, Madingley Road, Cambridge CB3 0HA, UK\\
$^{9}$Department of Astronomy, University of California Berkeley, Berkeley, CA 94720-3411, USA\\
$^{10}$ Center for Astrophysics ${\rm \mid}$ Harvard {\rm \&} Smithsonian, 60 Garden Street, Cambridge, MA 02138, USA \\ 
$^{11}$INAF -- Osservatorio Astrofisico di Torino, via Osservatorio 20, IT-10025 Pino Torinese, Italy\\
$^{12}$Observatoire Astronomique de l'Universit\'e de Gen\`eve, Chemin Pegasi 51b, CH-1290 Versoix, Switzerland\\
$^{13}$INAF -- Fundaci\'on Galileo Galilei, Rambla Jos\'e Ana Fernandez P\'erez 7, ES-38712 Bre\~na Baja, TF, Spain\\
$^{14}$INAF -- Osservatorio Astronomico di Palermo, P.za Parlamento 1, IT-90134 Palermo, Italy\\
$^{15}$INAF -- Osservatorio Astronomico di Cagliari, via della Scienza 5, IT-09047 Selargius, Italy\\
$^{16}$INAF -- Osservatorio Astronomico di Brera, via E.~Bianchi 46, IT-23807 Merate (LC), Italy\\
$^{17}$Centre for Exoplanets and Habitability, University of Warwick, Coventry CV4 7AL, UK\\
$^{18}$Department of Physics, University of Warwick, Coventry CV4 7AL, UK\\
$^{19}$Astrophysics Research Centre, School of Mathematics and Physics, Queen's University Belfast, Belfast BT7 1NN, UK
}

\date{Accepted 2021 July 26. Received 2021 July 12; in original form 2021 May 29}

\pubyear{2021}

\begin{document}
\label{firstpage}
\pagerange{\pageref{firstpage}--\pageref{lastpage}}
\maketitle

\begin{abstract}
To date, only 18 exoplanets with radial velocity (RV) semi-amplitude $<2$~\mps\ have had their masses directly constrained. The biggest obstacle to RV detection of such exoplanets is variability intrinsic to stars themselves, e.g.\ nuisance signals arising from surface magnetic activity such as rotating spots and plages, which can drown out or even mimic planetary RV signals. We use \mystar\ -- known to host three transiting planets, one of which, Kepler-37d, should be on the cusp of RV detectability with modern spectrographs -- \hl{as a case study in disentangling planetary and stellar activity signals. We show how two different statistical techniques -- one seeking to identify activity signals in stellar spectra, and another to model activity signals in extracted RVs and activity indicators} -- can each enable a detection of the hitherto elusive Kepler-37d. Moreover, we show that these two approaches \hl{can be} complementary, and in combination, facilitate a definitive detection and precise characterisation of Kepler-37d. Its RV semi-amplitude of $1.22\pm0.31$~\mps\ (mass $5.4\pm1.4$~$M_\oplus$) is formally consistent with TOI-178b's $1.05^{+0.25}_{-0.30}$~\mps, the latter being the smallest detected RV signal of any transiting planet to date, though dynamical simulations suggest Kepler-37d's mass may be on the lower end of our $1\sigma$ credible interval. Its consequent density is consistent with either a water-world or that of a gaseous envelope ($\sim0.4\%$ by mass) surrounding a rocky core. Based on RV modelling and a re-analysis of Kepler-37 TTVs, we also suggest that the putative (non-transiting) planet Kepler-37e should be stripped of its `confirmed' status.
\end{abstract}


%%%%%%%%%%%%%%%%%%%%%%%%%%%%%%%%%%%%%%%%%%%%%%%%%%

%%%%%%%%%%%%%%%%% BODY OF PAPER %%%%%%%%%%%%%%%%%%
\section{Introduction}\label{sec:intro}

%[Note --- things still \todo{to be added}/\tocheck{checked before submission}.]

Since the discovery of the first exoplanet over 25 years ago \citep{mayor1995}, Doppler spectroscopy -- also known as the radial velocity (RV) method -- has been a cornerstone of exoplanetary science. While it is a key tool for planet discovery, it is also indispensable for confirming and characterizing candidates discovered via other techniques, particularly transit photometry \citep[e.g.][]{konacki03}. As the RV method constrains planetary masses, it can shed light on planets' likely compositions, formation histories, atmosphere scale heights, and more. As of \tocheck{May 2021}, % FIX 
Doppler spectroscopy has been responsible for the discovery of around one in five known exoplanets, while Doppler spectroscopy and transit photometry have, together, led to the discovery of around $95\%$ of all confirmed exoplanets.\footnote{Based on counts from the NASA Exoplanet Archive, available online at \url{exoplanetarchive.ipac.caltech.edu}.} 

The precision of RV surveys has been steadily improving over the years, thanks to numerous technical advances. While the RV spectrographs of fifty years ago produced RVs with nominal errors in excess of $1$~\kmps\ per measurement, absorption-cell spectrographs \citep{cwalker79,marcybutler92,butler1996} have in recent years demonstrated precisions of order $1$~\mps\ \citep{butler2017}, whereas the newest generation of ultra-stabilized spectrographs \citep[i.e.\ using the so-called `simultaneous reference' technique:][]{Bara96,probst2014} today boast sub-\mps\ precisions, and aim to achieve $10$~\cmps\ precisions \citep{probst2014,HARPS3,schwab2016,petersburg20,pepe2021}. Such precision is sufficient, in principle, to detect the signal of an Earth-analogue exoplanet. Ambitious plans for next-generation ultra-stabilized spectrographs call for stability at the $1$~\cmps\ level \citep{pasquini2008,fischer2016eprv}. % TODO: update references

Despite enormous advances in instrumentation, a few significant obstacles impede the discovery or characterisation of planets with RV signatures at the $\lesssim1$~\mps\ level. The semi-amplitude $K_{\rm p}$ of the RV reflex motion induced by an orbiting planet scales as $K_{\rm p}\propto P_{\rm p}^{-1/3}$ and $K_{\rm p}\propto M_{\rm p} \sin i_{\rm p} \left( M_\star+M_{\rm p} \right)^{-2/3}$ \citep{perryman2011}, where $P_{\rm p}$ is the planet's orbital period, $M_{\rm p}$ is its mass, $i_{\rm p}$ is its angle of orbital inclination, and $M_\star$ is the mass of its host star. The paucity of $K_{\rm p} \lesssim 1$~\mps\ RV detections thus translates into a dearth of low-mass and, to a lesser extent, long-period RV exoplanets. 

\begin{figure*}
	\includegraphics[width=\textwidth]{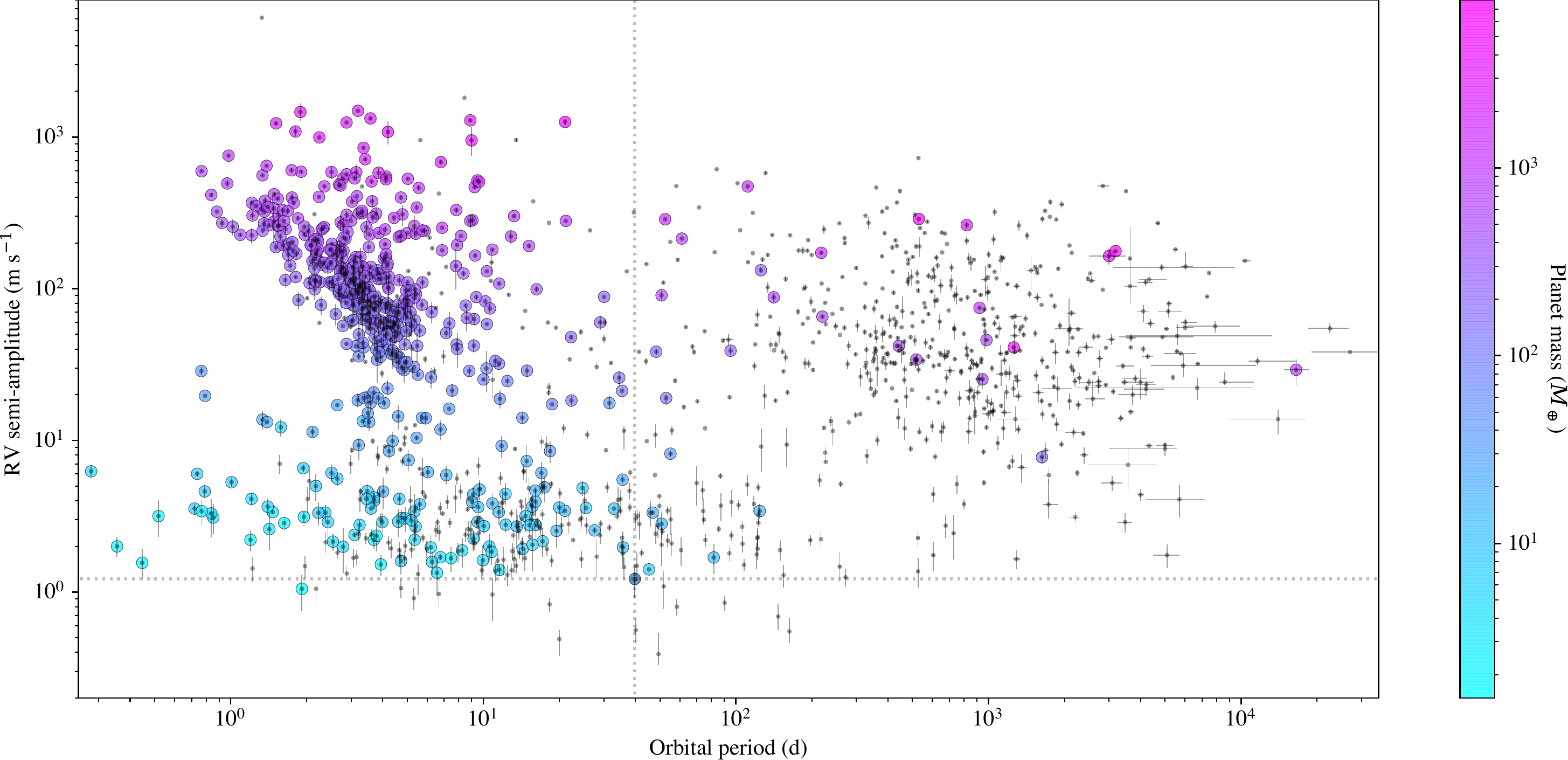}
    \caption{RV semi-amplitudes vs.\ orbital periods of the \tocheck{$1321$ (at the time of writing)} exoplanets with at least a $3\sigma$ detection in RVs. Of these, \tocheck{$556$ have known masses} (rather than minimum masses), which are indicated via the colour scale. All data retrieved from the NASA Exoplanet Archive on 2021 May 4, except for the point corresponding to Kepler-37d -- indicated by the intersection of the grey, dotted lines -- which is characterised in this work.}
    \label{fig:exo-archive}
\end{figure*}

At the time of writing, the NASA Exoplanet Archive contained \tocheck{1321} confirmed exoplanets with RV signals detected and inconsistent with zero at a $3\sigma$ level; of these, only 556 have a true mass (rather than minimum mass $M_{\rm p} \sin i_{\rm p}$) measurement. Moreover, of the latter 556 planets, only 18 have RV semi-amplitudes $<2$~\mps, with the smallest being TOI-178b's $1.05^{+0.25}_{-0.30}$~\mps \citep{leleu21}: still an order of magnitude larger than the precision of ESPRESSO, the spectrograph that characterised it \citep{pepe2021}, or of other cutting-edge instruments such as EXPRES \citep[e.g.][]{blackman20,petersburg20}. See Fig.~\ref{fig:exo-archive}.

The most vexatious obstacle to $\lesssim1$~\mps\ detections -- and thus the detection of Earth analogues -- is variability intrinsic to stars themselves. These stellar nuisance signals, due e.g.\ to surface magnetic activity such as rotating spots and plages, may be characterised by covariance structures similar to -- but amplitudes orders of magnitudes larger than -- signals expected from orbiting exoplanets \citep{dumusque2012}. Thus they can drown out or even mimic exoplanetary signals. 

There has therefore been significant effort devoted to developing ways to disentangle stellar activity signals from planetary ones \citep[][]{boi+09,lan+10,aigrain2012,tuomi2014,haywood2014b,robertson2014,Rajpaul2015,jones2017,gilbertson20}. These approaches have been modestly successful, e.g.\ enabling the characterisation of planets that would otherwise have remained undetected. Up until quite recently, though, most such efforts have been based on \emph{post hoc} attempts to model stellar activity signals already present in RVs. That is, RVs produced by a given pipeline are taken as a starting point, then combined with supplementary information (e.g.\ knowledge of a stellar rotation period; ancillary photometry; or spectroscopic diagnostics that should be sensitive to activity but not planets) to assess which variability in RVs might be attributable to stellar variability and which to planets. 
 
What about the process of getting actual RV measurements out of stellar spectra, however? For several decades, the standard approach to extracting RVs from spectra taken with stabilized spectrographs was to align or cross-correlate observed spectra with a template \citep{griffin67,simkin74,baranne79,tonry79,bouchy01}. The latter is typically either a synthetic template based on model stellar atmospheres, knowledge of atomic line locations, etc., or a high signal-to-noise ratio (SNR) spectrum derived from real observations \citep[e.g.][]{Nord94,Bara96,balona02}, in either case with various numerical weights and/or masking applied to different parts of the template \citep{pepe02}. The CCF approach retains wide currency and is employed, for example, in the primary data reduction pipelines of HARPS \citep{rupprecht04} and HARPS-N \citep{cosentino12}, as well as in the pipelines of newer, third-generation instruments such as ESPRESSO \citep{espresso18}.

More recently, though, there has been a rapidly growing number of efforts to develop improved ways of extracting RVs from stellar spectra, \hl{particularly with a view to mitigating stellar activity and/or telluric contamination from the final RVs} \citep[e.g.][]{anglada12,zech2018,dumusque18,wobble19,zhao20,Rajpaul2020,cc20,debeurs20}. These techniques are often data-driven, exploiting the vast quantity of wavelength- and/or time-dependent information in stellar spectra -- each typically containing hundreds of thousands of flux/wavelength pairs -- to extract `cleaner' RVs (\hl{i.e.~less contaminated by activity and/or telluric signals}) than may have been possible with traditional approaches. \hl{Some techniques} \hl{also employ forward modelling of observed stellar spectra in the process of deriving RVs} \citep[e.g.][]{zech2018,petersburg20}. 

In this paper, we use HARPS-N spectra of Kepler-37 as a case study in both stellar activity modelling (i.e.\ modelling stellar signals already present in RVs extracted by some pipeline) and \hl{spectral-level mitigation (identifying and suppressing the effects of activity at the level of stellar spectra,} in order to extract `cleaner' RVs). \hl{For short, throughout this paper we shall refer to these approaches simply as `modelling' and `mitigation'.}  We shall show how either approach can facilitate the detection and characterisation of a hitherto-elusive planet, and moreover, how stellar activity mitigation and modelling can be leveraged in tandem for superior results.

%While Kepler-37 is orbited by three exoplanets that have been observed to transit \citep{barclay+13}, the inner two planets lie well below the threshold of current-generation RV detectability. The third, however -- Kepler-37d, a $1.92$~$R_\oplus$ super-Earth with $39.79$~d orbital period -- could be on the cusp of RV detectability with a modern spectrograph such as HARPS-N. %We show how (i) stellar activity modelling, or (ii) stellar activity mitigation, in either case using state-of-the-art tools, can by itself enable a RV detection of the otherwise elusive Kepler-37d. Moreover, we demonstrate that these two approaches are complementary, and in combination lead to a definitive detection and a more precise characterisation of Kepler-37d. \tocheck{We thus characterise Kepler-37d as the transiting planet with the second smallest detected RV signal to date.}

The remainder of the paper is structured as follows. The next section draws on extant literature to summarise salient information about Kepler-37 and its planetary system. Section~\ref{sec:HN-obs} provides information about the HARPS-N observations that form the basis of this study, and we use these observations to derive updated stellar parameters for Kepler-37. Section~\ref{sec:methods1} describes our methods for extracting RVs and mitigating activity when doing so, while Section~\ref{sec:methods2} discusses our modelling of extracted RVs. We present and discuss our main results in Section~\ref{sec:results}, then conclude in Section~\ref{sec:discuss}. Additionally, Appendix~\ref{sec:HIRES} contains comments on supplementary (though ultimately inconclusive) analyses we performed using HIRES spectra.

\section{The Kepler-37 system}\label{sec:K37-lit}

\subsection{The star}

Kepler-37 -- also variously designated TYC~3131-1199-1, KIC8478994, KOI~245, and UGA-1785 -- is a main sequence dwarf, slightly cooler and smaller than the Sun, situated $64$~pc from Earth in the Lyra constellation. Table~\ref{tab:K37-star} summarises various key properties. Detailed abundances for 19 elements, derived using Keck/HIRES spectra, are provided in \citet{schuler15}.

Though its spectral type has not been formally determined \citep{schuler15}, its temperature, luminosity, mass, radius, etc.\ are indicative of a late-G or early-K dwarf \citep[G8V/K0V:][]{pecaut13}. Despite the low luminosity and low amplitude oscillations associated with cool dwarfs, \citet{barclay+13} were able to detect asteroseismic oscillations of Kepler-37 in Kepler photometry: at the time of the aforesaid analysis, Kepler-37 was the smallest and densest star in which Solar-like oscillations had been detected \citep[see also][]{huber13}, though it no longer holds this record.

Kepler-37 is undetected in the ROSAT All Sky Survey \citep{rosat16}, indicating that it is a weak X-Ray source, which would be consistent with low activity levels. Nevertheless, given Kepler-37's probable spectral type, it should have an outer convective envelope; given also its estimated $\sim4$~Gyr age and $\sim29$~d rotation period, one could expect that rotational modulation of possible stellar magnetic features may lead to non-trivial photometric and (apparent) RV variability \citep{schirjver2000}. This inference will be borne out in Section~\ref{sec:methods1}, where a cursory examination of our Kepler-37 RVs will reveal significant correlations with stellar activity indicators.

% https://iopscience.iop.org/article/10.1088/0004-637X/815/1/5/meta

\subsection{Planetary system}

\subsubsection{Three transiting planets}

\mystar\ is orbited by three known transiting planets, the periods of which are within one per cent of a 5:8:15 commensurability, hinting at a possible mean-motion resonance chain \citep{barclay+13}.

%At the time its discovery was announced in 2013, the inner planet had the smallest radius of any known transiting planet. 
With a radius of just $0.30\,R_\oplus$, the inner planet, \mystar b, is smaller than Mercury, and only slightly larger than the Moon. The NASA Exoplanet Archive indicates that, as of May 2021, \mystar b still holds the distinction of having the smallest measured radius of any exoplanet. The next smallest transiting planet, Kepler-444b, has a radius of $0.40\,R_\oplus$, while only two other confirmed planets, {Kepler-102b} and \hbox{Kepler-444c}, have radii estimates smaller than $0.5\,R_\oplus$. \mystar b orbits \mystar\ with a period of $13.37$~d at a distance of $0.10$~AU, and is probably rocky, with no atmosphere or water \citep{barclay+13}.

\mystar c is around three-quarters of the size of Earth -- still very much at the lower end of the radius distribution of confirmed exoplanets  -- and orbits \mystar\ every $21.30$~d at a distance of $0.14$~AU. \mystar d, on the other hand, is approximately twice the size of Earth, and orbits \mystar\ every $39.79$~d at a distance of $0.21$~AU. Given their proximity to \mystar, \mystar c and \mystar d are also not expected to contain surface liquid water.

Some key properties of the three transiting planets are summarised in Table~\ref{tab:planet-properties}.

\subsubsection{Kepler-37e?}\label{sec:k37e?}

The paper presenting the discovery of the three transiting planets \citep{barclay+13} noted the presence of a fourth planet candidate orbiting \mystar, then referred to as KOI-245.04, with a $51.2$~d orbital period. However, \citet{barclay+13} commented that they did `not trust that KOI-245.04 is a valid planet candidate' as the inclusion of more data since the release of the (then) most recent planet candidate catalogue \emph{decreased} the signal to noise of the putative transit signal, suggesting the transit-like signal in question was likely caused by random noise, or correlated stellar or instrumental noise.

Subsequently, \citet{hadden+14} analysed TTVs obtained \hl{over three years (Q1--Q12)} of the Kepler mission, to extract densities and eccentricities of 139 sub-Jovian planets, using ephemerides from the TTV catalog published by \citet{mazeh13}. In the former authors' analysis, the apparent TTVs of `\mystar e' were considered to derive constraints on the mass of \mystar d; however, this seemed to pre-suppose that KOI-245.04 was a real planet, which (to the best of our knowledge) had not yet been established. 

\hl{A number of later studies suggested} that Kepler-37's transiting planets do \emph{not} exhibit significant TTVs \citep[e.g.][]{holczer16,Gajados19}. Sowing further doubt, two separate planet-searching pipelines \citep{huang14,kunimoto20} applied to Kepler data explicitly failed to identify the signal associated with KOI-245.04 as a \emph{bona fide} transit. In fact, \citet{kunimoto20} noted that KOI-245.04 was the only `planet' completely missed by the pipeline they applied to $\sim200\,000$ FGK dwarfs observed by Kepler.

Given the \hl{scant evidence in support of a detection, and the lack of a literature consensus}, we hereafter tread with caution and treat KOI-245.04 or `Kepler-37e' as a \emph{putative} but not indisputably confirmed planet, despite several catalogs treating it as such.\footnote{\hl{At the time of writing, `Kepler-37e' is listed as a confirmed planet by sources including} the NASA Exoplanet Archive (in which Kepler-37e bears no `controversial' flag), the Open Exoplanet Catalog  (available online at \url{www.openexoplanetcatalogue.com}), and SIMBAD \citep{SIMBAD}. \hl{The Extrasolar Planets Encyclopaedia} \citep{exo-encyc} \hl{dissents, listing only three confirmed planets in the system. 
}} We shall aim to shed light on this candidate planet's existence through our main RV analysis, complemented by an updated TTV investigation.

\begin{table}
\centering
    \caption{A few key properties of \mystar, drawn from the following sources: (1) \citet{Stassun17}; (2) \citet{Berger18}; (3) \citet{schuler15}; (4) \citet{morton16}; (5) \citet{gaia_mission}; (6) \citet{gaia_dr2_summary}; (7) \citet{apsis18}; (8) \citet{gaia-edr3}; (9) \citet{Walkowicz13}; (10) \citet{tycho2}; (11) \citet{tycho2_johnson}; (12) \citet{tycho2_johnson2}; (13) \citet{2mass_catalog}; (14) \citet{cutri14}.}
    \begin{tabular}{l c c}
    \hline
    Stellar property & Value & Reference\\ \hline
        Mass $(M_	\odot)$ & $0.87 \pm 0.15$ & (1) \\
        Radius $(R_\odot)$ & $0.787^{+0.033}_{-0.031}$ & (2) \\
        Density ($\rho_\odot$) & $1.76\pm0.23$ & (1) \\
        T$_{\rm{eff}}$ (K) & $5406 \pm 28$  & (3) \\
        $\log g$ (\cmpss) & $4.49\pm0.13$ & (3) \\
        Age (Gyr) & $3.8^{+3.3}_{-2.0}$ & (4) \\
        $[\rm{Fe}/\rm{H}]$ (dex) & $-0.32 \pm 0.07$ & (2)  \\
        Luminosity ($L_\odot$) & $0.4791\pm0.0014$ & (5), (6), (7) \\
        Distance (pc) & $63.999 \pm 0.043$ & (5), (6) \\
        Spectroscopic $P_{\rm rot}$ (d) & $22.9\pm6.9$ & (9) \\
        Photometric $P_{\rm rot}$ (d) & $28.8\pm3.3$ & (9) \\
        $v \sin i$ (\kmps) & $1.70\pm0.50$ & (9) \\
        RA (J2000) & $284.0592\pm0.0088\degr$ & (5), (8) \\
        Dec (J2000) & $44.518\pm0.011\degr$ & (5), (8) \\
        Parallax (mas) & $15.625\pm0.010$ & (5), (8) \\
        Absolute RV (\kmps) & $-30.58\pm0.30$ & (5), (6) \\ 
        B-band (mag) & $10.446\pm0.028$ & (10), (11), (12) \\
        V-band (mag) & $9.770 \pm 0.028$ & (10), (11), (12) \\
        J-band (mag) & $8.356\pm0.018$ & (13) \\
        H-band (mag) & $8.000\pm0.024$ & (13) \\
        K-band (mag) & $7.942\pm0.013$ & (13) \\
        W1 (mag) & $7.867\pm0.025 $ & (14) \\
        W2 (mag) & $7.933\pm0.020$ & (14) \\
        W3 (mag) & $7.901\pm0.019$ & (14) \\

    \hline
    \label{tab:K37-star}
    \end{tabular}
\end{table}

\begin{table*}
	\centering
	\caption{Key properties of the three planets known to transit Kepler-37, and of the putative fourth planet Kepler-37e. Data from the following sources: (1) \citet{Gajados19}; (2) \citet{Berger18}; (3) \citet{Stassun17}; (4) \citet{vaneylen+15}; (5) \citet{barclay+13}; (6) \citet{hadden+14}.}
	\label{tab:planet-properties}
 \begin{tabular}{ccccccccccc}
 \hline
 Planet  & Orbital period  & Ref.  & Radius  & Ref. & Inclination  & Ref.  & Eccentricity & Ref. & Transit mid-point & Ref.\ \\ 
 & (d) & & ($R_\oplus$)  & & (deg) & & & & (BJD$-$2\,455\,000) & \\
 \hline
 Kepler-37b  & $13.367020\pm0.000060$  & (1)  & $0.277^{+0.033}_{-0.025}$ & (2) & $88.63\pm0.41$  & (3)  & $0.08^{+0.210}_{-0.080}$ & (4) &$17.0473\pm0.0037$ & (1) \\
 Kepler-37c  & $21.301848\pm0.000018$  & (1)  & $0.725^{+0.040}_{-0.032}$  & (2) & $89.07^{+0.19}_{-0.33}$ & (5)  & $0.090^{+0.180}_{-0.090}$ & (4) & $24.839 97 \pm 0.000 87$ &  (1) \\
 Kepler-37d  & $39.7922622\pm0.0000065$  & (1)  & $1.917^{+0.085}_{-0.076}$ & (2) & $89.335^{+0.043}_{-0.047}$ & (5)  & $0.15^{+0.07}_{-0.10}$ & (4) & $8.24982 \pm 0.00013$ & (1) \\
 `Kepler-37e'  & $\sim51.196$? & (6)  & ? & --- & ? & --- & ? & --- & n/a & --- \\ \hline
 \end{tabular}

\end{table*}

\subsubsection{Expected RV detectability}\label{sec:planet-detectability}

Using our knowledge of the radii of the transiting planets, we can calculate the semi-amplitude of their associated RV signals, assuming various possible compositions \citep{zeng19}.

Even if \mystar b had \hl{an improbably-high density, it would certainly \emph{not} be detectable by HARPS-N, nor indeed by any extant spectrograph}.  Assuming a $100\%$~Fe composition, say, its consequent $0.03~M_\oplus$ mass would induce reflex motion in \mystar\ with $\lesssim1$~\cmps\ semi-amplitude, which is an order of magnitude smaller than the semi-amplitude of the RV reflex motion induced in the Sun by Earth. Kepler-37c has equally unrealistic prospects for RV characterisation: a $100\%$~Fe composition would give rise to a meagre $14$~\cmps\ RV signal.

Kepler-37d, on the other hand -- the focus of this study -- has more realistic prospects for RV detection. \hl{Considering all planets in the NASA Exoplanet Archive with radii within $0.25$~$R_\oplus$ of Kepler-37d's radius (and non-trivial mass measurements), $84\%$ have masses greater than $4.1$~$M_\oplus$, and $50\%$ have masses greater than $7.3$~$M_\oplus$; assuming the latter masses for Kepler-37d, its consequent RV signal would be $0.9$~\mps~or $1.6$~\mps, respectively.} Earth-like rocky ($32.5\%$~Fe and $67.5\%$~$\rm{MgSiO}_3$) and $100\%$~$\rm{MgSiO}_3$ compositions for Kepler-37d would be associated with RV signals with semi-amplitudes of $2.7$~\mps\ and $1.9$~\mps, respectively. Even a $50\%$ Earth-like rocky core plus $50\%$ mass in an H$_2$O layer, assuming $500$~K equilibrium temperature and 1~mbar surface pressure, would be associated with a $1.1$~\mps\ RV signal: difficult, though not unfeasible, to detect with HARPS-N (cf.\ remarks in Section~\ref{sec:intro} on the dearth of $<2$~\mps\ detections). 

\section{HARPS-N observations}\label{sec:HN-obs}

\subsection{Details of observations}\label{sec:obs-detail}

The basis of the analyses in this paper is a set of 110 high-SNR spectra obtained with HARPS-N \citep{cosentino12}, an $R\approx115\,000$ spectrograph covering a wavelength range from 383 to 690~nm.

A total of 115 HARPS-N spectra were obtained between 2014 April 15 and 2019 September 10, with the majority from 2014 and 2015 (59 and 53 spectra, respectively), and only three spectra taken in 2019. The majority of the observing time originated from two proposals in period AOT29 and AOT31 (PI Buchhave) that were respectively awarded 29.5 and 17.5~hr of observing time for Kepler-37, corresponding to roughly 94 spectra. The remaining observations were acquired by the HARPS-N GTO program. 108 of the 115 spectra had exposure times of $1800$~s, with the remaining exposure times being between $930$~s and $1500$~s. Across all observations, \hl{the median SNR/pixel at $570$~nm was 115;} the median seeing was $1.27''$, and the median airmass was 1.16.

Following visual inspection of RVs extracted by the HARPS-N Data Reduction Software (DRS, see Section~\ref{sec:DRS}), we identified five RVs as possible outliers, viz.\ from spectra taken on the nights of 2014 April 20, 2014 July 11, and 2015 May 11, 14, and 15. The SNRs for all five spectra were several (up to about ten) times lower than for the majority of spectra. Moreover, some of these spectra were associated with activity indicator values that appeared more anomalous than their formal error bars suggested (e.g.\ \lrhk), DRS drift correction quality control flags marked as `failed', unusually poor seeing, etc. To err on the side of caution, we excluded these five spectra from further consideration, and worked with the remaining 110 spectra. 

\subsection{Refined stellar parameters}

\begin{table}
\centering
    \caption{Refined properties of \mystar\ derived in this work using our HARPS-N spectra. Notes: (1) averaged parameters from ARES$+$MOOG, \texttt{CCFPams}, and SPC analyses; (2) microturbulent velocity -- from  ARES$+$MOOG analysis; (3) from SPC analysis; (4) averaged parameters from \texttt{isochrones} and MIST analyses.}
    \label{tab:K37-star2}
    \begin{tabular}{l c c}
    \hline
    Stellar property & Value & Note\\ \hline
        %Spectral type & ???  & ??? \\ % TODO: CHECK
        T$_{\rm{eff}}$ (K) & $5357 \pm 68$  & (1) \\
        $\log g$ (\cmpss) & $4.60\pm0.12$ & (1) \\
        $[\rm{Fe}/\rm{H}]$ (dex) & $-0.36 \pm 0.05$ & (1)  \\
        $\xi_t$ (\kmps) & $0.93\pm0.08$ & (2)\\
        $v \sin i$ (\kmps) & $<2.0$ & (3) \\
        Mass $(M_	\odot)$ & $0.790_{-0.030}^{+0.033}$ &  (4) \\
        Radius $(R_\odot)$ & $0.789^{+0.0064}_{-0.0056}$ & (4) \\
        Density $(\rho_\odot)$ & $1.624^{+0.096}_{-0.093}$ & (4) \\
        Age (Gyr) & $7.6^{+3.4}_{-3.1}$ & (4) \\
        Distance (pc) & $63.999\pm0.042$ & (4) \\

    \hline
    \end{tabular}
\end{table}

The high SNR needed to extract precise RVs from spectra means spectra used for this purpose are generally more than adequate for deriving stellar parameters. Accordingly, we took advantage of our high SNR, high-resolution HARPS-N spectra to derive refined stellar parameters for \mystar. 

We derived stellar atmospheric parameters via three independent methods: ARES$+$MOOG, \texttt{CCFPams}, and the Stellar Parameter Classification tool (SPC). ARES$+$MOOG is a curve-of-growth method based on neutral and ionized iron lines, and is explained in \citet{sousa14} and references therein; the \texttt{CCFPams} method, described in \citet{malavolta17}, uses cross-correlation function (CCF) equivalent widths to obtain effective temperature, surface gravity, and iron abundance via empirical calibration; lastly, SPC is a spectrum synthesis method, described in detail in \citet{buchhave12} and \citet{buchhave14}. The first two methods use a co-added master spectrum, while SPC uses the individual spectra and takes a median of the individual results. Surface gravity estimates from ARES$+$MOOG and \texttt{CCFPams} were corrected for accuracy \citep{mortier14}, and systematic errors were added to our precision errors for effective temperature, surface gravity, and iron abundance as derived by ARES$+$MOOG and \texttt{CCFPams} \citep{sousa11}. Our final adopted stellar atmospheric parameters, which appear in Table~\ref{tab:K37-star2}, are inverse variance-weighted averages of the results from these three methods, following a methodology first used by \citet{malavolta18} that has since been well tested and widely adopted  \citep[e.g.][]{rice19,mortier20}.

We obtained stellar mass, radius, age and distance estimates from isochrones and evolutionary tracks. We used the \texttt{isochrones} package \citep{morton15} and two separate stellar evolution models, viz.\ Dartmouth \citep{dotter08} and MIST \citep[\texttt{MESA} Isochrones and Stellar Tracks -- ][]{dotter16} to estimate these parameters, following the methodology described in \citet{mortier20}. As inputs, we used our spectroscopically-determined effective temperature and iron abundance, the Gaia EDR3 parallax, and magnitudes from the visible to mid-infrared (as listed in Table~\ref{tab:K37-star}). We ran each code three times -- varying only between the sets of spectroscopic parameters estimates by the ARES$+$MOOG, \texttt{CCFPams} and SPC approaches -- to produce a total of six estimates for stellar mass, radius, age and distance. Our final estimates for each parameter, in Table~\ref{tab:K37-star2}, were obtained by combining the six posterior distributions for each parameter.

We note that all of our revised stellar parameters in Table~\ref{tab:K37-star2} are consistent within $1\sigma$ with the literature values in Table~\ref{tab:K37-star}. While the precisions of our surface gravity and metallicity estimates are essentially the same as the literature values, our stellar mass and radius estimates have error bars five times smaller than the literature values (thanks to a much improved parallax value). Furthermore, while the parameters in Table~\ref{tab:K37-star} are drawn from many distinct data sets and analyses, our stellar parameters have the advantage of being derived in a uniform way from a single data set.

\section{RV extraction}\label{sec:methods1}

We consider RVs extracted from HARPS-N spectra using two independent methods: the HARPS/HARPS-N DRS pipeline, and a Gaussian process (GP) method that infers RVs by aligning pairs of spectra. 

\subsection{DRS RV extraction and activity indicators}\label{sec:DRS}

Both HARPS and HARPS-N spectra are, by default, processed by the DRS pipeline, which is optimised for exoplanet searches \citep{lovis06,cosentino12}. The DRS cross-correlates observed spectra with a stellar mask chosen from a template library, and thus computes a CCF for each spectrum. The mask itself consists of a list of wavelength ranges that identify spectral lines, and weights defining the contribution of each spectral line to the cross-correlation. Masks optimised for G2, K5 and M2-type stars are readily available and have been widely used, though more recently, bespoke masks for other stellar types have also been developed \citep[e.g.][]{rainer20}. The DRS uses the resulting CCF both for RV extraction and for building stellar activity indicators, since e.g.\ stellar activity can induce asymmetries and other distortions to the CCF. 

\begin{figure*}
	\includegraphics[width=\textwidth]{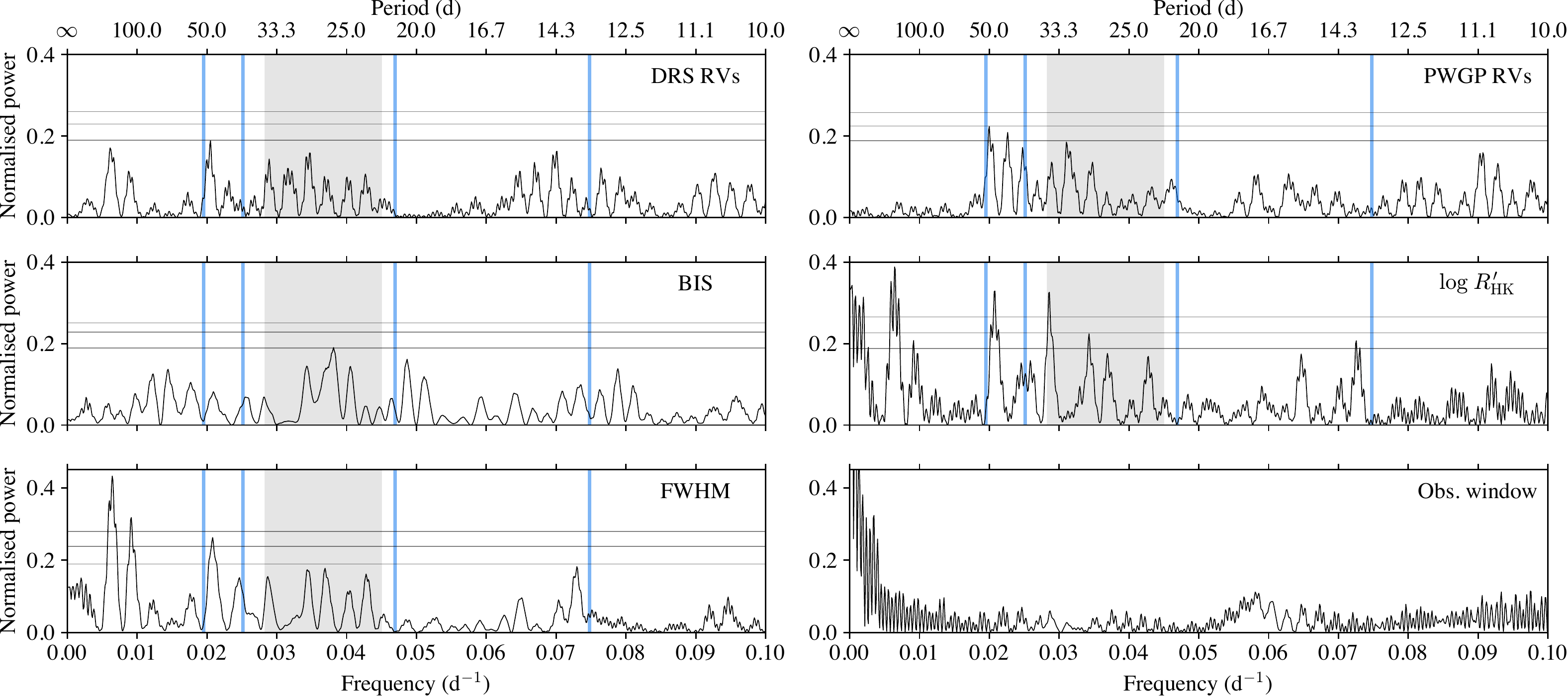}
    \caption{GLS periodograms of Kepler-37 RVs, as extracted by the DRS and PWGP approaches (top panels); of three different stellar activity indicators (middle and lower left panels); and of the observing times (lower right panel). Vertical blue lines indicate the orbital periods of Kepler-37b through to Kepler-37e, while the grey shaded box covers a $\pm2\sigma$ credible interval around Kepler-37's photometric $P_{\rm rot}$, as estimated by \citet{Walkowicz13}. The horizontal lines indicate, from bottom to top, estimated $10\%$, $1\%$, and $0.1\%$ false alarm probability (FAP) thresholds, respectively.}
    \label{fig:K37_Lomb_data}
\end{figure*}

\begin{figure*}
\centering
	\includegraphics[width=\textwidth]{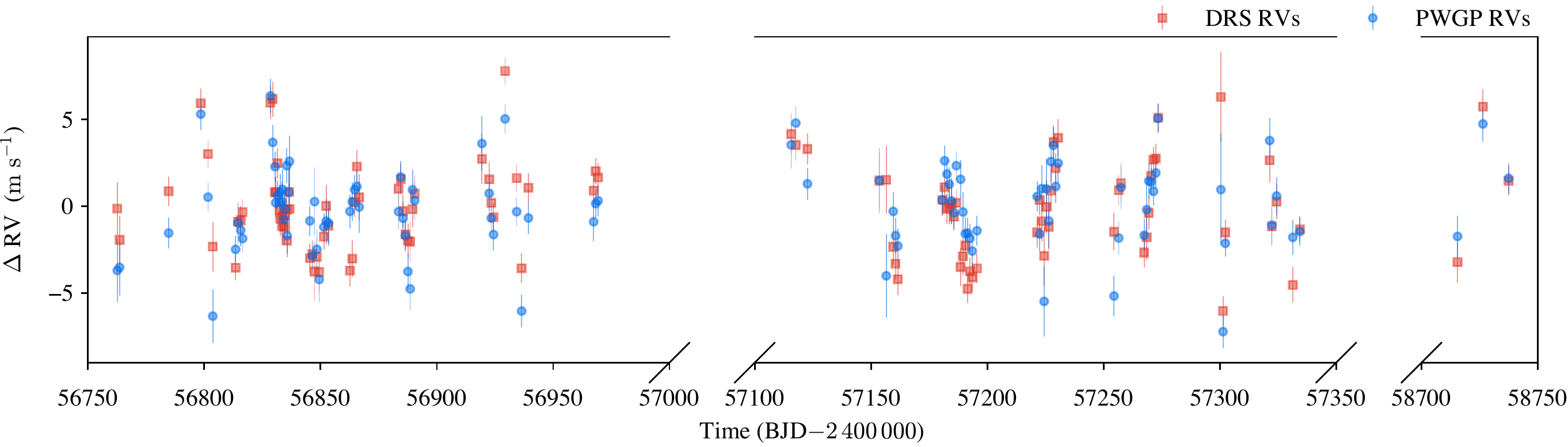}
    \caption{The 110 Kepler-37 HARPS-N RVs used in this paper, as extracted by the DRS pipeline (red squares) and PWGP method (blue circles), along with $1\sigma$ error bars from each method. The DRS RVs have their mean value of $-30.651$~\kmps\ subtracted, to facilitate comparison with the PWGP RVs, which by construction have zero mean velocity.}
    \label{fig:RVs}
\end{figure*}

As the DRS has been widely used in published RV studies, we do not discuss it in  detail here; we simply note that we used a K5 mask and version 3.7 of the DRS -- the latest version available in 2020, when most of our analyses were carried out -- to extract our RVs.\footnote{As of 2021, a newer version of the DRS (counter-intuitively numbered 2.2.8), based on the ESPRESSO pipeline, is available. The difference in \mystar\ RVs extracted by the two DRS versions turns out to be negligible: $95\%$ of RVs are consistent within $2\sigma$, and $100\%$ within $2.3\sigma$. Correlations between RVs and activity indicators (see Table~\ref{tab:RV_correlations}) are as strong or marginally stronger in the newer RVs. We also re-ran a representative subset of our modelling (per Section~\ref{sec:methods2}) using the newer RVs; our conclusions were identical.} \hl{A periodogram and time series representation of these RVs appear in Figs}~\ref{fig:K37_Lomb_data} \hl{and} \ref{fig:RVs}, \hl{respectively.} In addition to RVs, we also extracted activity indicators including \lrhk, bisector span (\bis), and FWHM time series; the activity indicators referred to in the rest of this paper are always those extracted by the DRS, since our second method of RV extraction, described below, does not compute these activity indicators. 

Summary statistics pertaining to key stellar activity indicators for \mystar, derived by the DRS, are given in Table \ref{tab:data-summary}. In particular, we note here that the mean \lrhk\ value over our $110$ spectra, $-4.871$, would lead Kepler-37 to be classified as (relatively) `inactive,' following the scheme given by \citet{maldo10}, in which a star is classified as `active' when $\lrhk>-4.75$, and `very active' when $\lrhk>-4.2$. For comparison, the Sun varies between about $\lrhk\sim-4.85$ and $-5.0$ as it moves between high- and low-activity phases. 

Generalised Lomb-Scargle \citep[GLS,][]{zechmeister2009} periodograms for the same stellar activity indicators appear in the middle panel of Fig.~\ref{fig:K37_Lomb_data}. Periodograms for the tightly-correlated \lrhk\ and FWHM time series show prominent peaks around $14$~d, $29$~d, $35$~d and $48$~d, all of which could be associated with \mystar's estimated $\sim29$~d rotation period (see Table~\ref{tab:K37-star}) and possible active region evolution. Power around $\sim100$~d and $\sim200$~d could be associated with active region evolution or changes as a result of a longer-term magnetic cycle \hl{(we touch again on the latter possibility in Section}~\ref{sec:param-posterior}). The BIS periodogram has a prominent peak around $26$~d, again close to the estimated rotation period. Note that the stellar rotation period should not be expected to be identifiable via a single periodogram peak \citep{nava2020}: the GLS periodogram assumes a data model of a \emph{single sine wave plus white Gaussian noise} \citep{vanderplas18}. This will always be an imperfect -- and sometimes deeply flawed -- model for quasi-periodic RVs containing multiple planets and signals arising from dynamic active regions on a diferentially-rotating star, which in turn might be subject to long-term magnetic cycles, etc. 

\subsection{Pairwise GP extraction}

\begin{table}
\centering
\caption{A few statistics summarising key properties of our HARPS-N spectra of Kepler-37, and measurements extracted from these spectra. To avoid confusion from assigning multiple meanings to the Greek letter sigma, in this table we use `SD' to denote the standard deviation of a given set of measurements, and `error' to refer to the estimated $1\sigma$ error bars on measurements.}
\label{tab:data-summary}
\begin{tabular}{lc}
\hline
Summary statistic            & Value  \\ 
\hline
No. spectra including outliers & 115    \\
No. spectra analysed in this work         & 110    \\
Time span of observations (d)                & 1974.8 \\
\hline
Mean (DRS RV) (\mps)               & $-30\,651.51$      \\
Mean (PWGP RV)  (\mps)            & $0$      \\
Mean (DRS RV error) (\mps)              & $1.02$      \\
Mean (PWGP RV error) (\mps)            & $1.11$      \\
SD (DRS RV) (\mps)                 & $2.68$      \\
SD (PWGP RV) (\mps)               & $2.50$      \\
\hline
Mean $\left( \rm{BIS} \right)$ (\mps)                   & $4.61$      \\
Mean (BIS error) (\mps)& 2.04 \\
SD (BIS) (\mps)                    & $3.47$      \\
\hline
Mean (\lrhk)                  & $-4.871$      \\
Mean (\lrhk\ error)                    & $0.0051$      \\
SD (\lrhk)                    & $0.023$      \\
\hline
%(6026.601636363635, 8.969583503786845, 2.401059090909091)
Mean (FWHM) (\mps)                  & $6026.60$      \\
Mean (FWHM error) (\mps)                    & $2.40$      \\
SD (FWHM) (\mps)                    & $8.97$      \\
\hline
\end{tabular}
\end{table}

\begin{table*}
	\caption{Non-parametric measures of correlation between \mystar\ RVs extracted using the DRS and the PWGP approach, and various time-varying quantities (columns) that are expected to be independent of dynamically-induced stellar velocity shifts, though possibly sensitive to stellar activity or telluric contamination. \hl{The FWHM, contrast and BIS are all CCF parameters yielded by the DRS; they measure the width, depth and asymmetry, respectively, of the average spectral line profile} \citep{anglada12}. BERV is the barycentric Earth radial velocity at the time of observation. For each correlation, we give the Spearman rank correlation coefficient, $\rho$, and the associated $p$-value. Correlations significantly nonzero at a $p<.05$ level are typeset in bold.}
	\label{tab:RV_correlations}
	\begin{tabular}{lcccccccccccc}
		\hline
		  & \multicolumn{2}{c}{FWHM} & \multicolumn{2}{c}{Contrast} & \multicolumn{2}{c}{BIS} & \multicolumn{2}{c}{\lrhk} & \multicolumn{2}{c}{Airmass} & \multicolumn{2}{c}{BERV}   \\
		 &$\rho$&$p$ &$\rho$&$p$ &$\rho$&$p$ &$\rho$&$p$ &$\rho$&$p$ &$\rho$&$p$ \\
		 \hline
		 
 		 DRS RVs &$\boldsymbol{+0.50}$&$\boldsymbol{\ll.001}$ &$\boldsymbol{-0.50}$&$\boldsymbol{\ll.001}$ &$\boldsymbol{+0.22}$&$\boldsymbol{.02}$ &$\boldsymbol{+0.45}$&$\boldsymbol{\ll.001}$ &${-0.03}$&$.76$ &$-0.12$&$.20$  \\
 		 
 		 PWGP RVs &${+0.12}$&${.23}$ &$-0.14$&$.15$ &$-0.04$&$.68$ &$+0.13$&$.19$ &$+0.01$&$.95$ &$-0.02$&$.85$  \\
 		 
		\hline
	\end{tabular}
\end{table*}

The second method we used to extract RVs was the GP-based method presented in \citet{Rajpaul2020}. In brief, GPs are used to model and align all pairs of spectra with each other; the pairwise RVs thus obtained are combined to produce differential stellar RVs, without constructing any template. Given the reliance on GP modelling of spectra, and the pairwise nature of the RV extraction, we hereafter refer to this method as the pairwise GP (PWGP) method. 

The rationale for the pairwise comparison of spectra is largely computational: modelling and aligning tens or hundreds of spectra simultaneously would require inversion of enormous covariance matrices (generally the biggest bottleneck to any GP modelling), whilst also sampling parameter spaces of very high dimensionality. By contrast, pairwise RV extraction entails relatively cheap, repeated computation that parallelises trivially, with only a few parameters to be optimised or sampled for any pair of spectra.

The PWGP method can be used to compute differential RVs on a localised basis, e.g.\ to yield an independent set of RVs for each \'echelle order, or for much smaller subdivisions of orders. The motivation is that regions of spectra affected by e.g.\ stellar activity or telluric contribution may be identified and excluded (essentially a data-driven masking of the spectrum, without any knowledge of line locations or properties) from the calculation of the final RVs, which are obtained by an inverse variance-weighted average of the localised RVs.

\citet{Rajpaul2020} showed that even a relatively crude implementation of the PWGP method, applied to an inactive star (where only modest if any improvements may have been expected compared to the DRS), resulted in RVs with precisions comparable to and rms scatter about $30\%$ lower than RVs extracted by the DRS and two other commonly-used codes, \emph{viz.} HARPS-TERRA \citep{anglada12} and SERVAL \citep{zech2018}. 

We applied the PWGP method essentially as described in the proof-of-concept paper by \citet{Rajpaul2020}. We divided each of the $69$ \'echelle orders per spectrum into $32$ `chunks,' \hl{for a total of $2208$ chunks,} each $128$ pixels wide (typical width a little under $2$~\AA). Corresponding chunks across all pairs of spectra were fitted with a Mat\'ern-$\tfrac{5}{2}$ kernel, and aligned via a maximum likelihood approach. This produced a $110\times110\times2208$ array of pairwise RV shifts, and an uncertainty array of identical dimension. The arrays are necessarily anti-symmetric and symmetric, respectively, with respect to the first two dimensions. %(the RV shift of spectrum $i$ vs $j$ is always the additive inverse of the shift of $j$ vs $i$) while the latter is symmetric with the respect to the same dimensions.

Up to this point, the PWGP method proceeds autonomously. However, some thought is then required as to how, if at all, to identify and exclude spectral chunks that are likely `contaminated' -- by stellar activity-related variability, as typified e.g.\ in the Calcium \textsc{ii} H \& K lines, by telluric variability, typified e.g.\ in water vapour lines, etc.\ -- before producing a final set of differential RVs. Here, approaches of varying levels of sophistication are possible. For example, a subset of local RVs exhibiting periodicities known to be associated with the stellar rotation period could be excluded, or a clustering analysis might help identify chunks with similar properties and link problematic chunks with one another. We adopted a conservative version of a simple approach that has at least been tested on both synthetic and real spectra \citep{Rajpaul2020,damasso20}; specifically, we excluded all chunks exhibiting the following simple properties:
\begin{enumerate}
    \item rms or median error bar $>10$~\kmps; and/or
    \item significant correlation ($p<.05$ under a non-parametric Spearman rank correlation test) with any activity indicator (BIS, \lrhk, FWHM), barycentric Earth RV, or airmass time series. 
\end{enumerate}
\hl{Assuming Poisson statistics, one would expect the SNR in RVs extracted from a given wavelength range to scale as $\sqrt{N}$, with $N$ being the photon count. All else being equal, and assuming a precision of $\sim1$~\mps~is possible with the full spectrum, a precision of $\sqrt{2208}\sim50$~\mps~might be expected from individual chunks}. Thus, local RVs satisfying condition (i), i.e.\  with rms scatter or error bars orders of magnitude larger than \hl{could have been expected \emph{a priori}}, were likely contaminated by extremely strong stellar, telluric or instrumental signals, or they may have corresponded to continuum-dominated regions of spectra containing very little Doppler information.\footnote{\hl{A single-value cut will inevitably be somewhat arbitrary. We note that we obtained nearly identical results when we used a $1$~\kmps~cut instead.}}

Local RVs satisfying condition (ii) likely suffered mild to strong stellar activity or telluric contamination. Note, however, that an absence of a significant local correlation with e.g.\ an activity indicator does not prove that a chunk was contamination free -- it simply means that any such contamination was not detected via one specific \hl{measure of statistical association}. In the case of very strong stellar variability, it can well be the case \citep[e.g.][]{damasso20} that even when combining local chunk RVs that all appear uncorrelated with activity indicators -- due to low SNR -- the resultant RVs nevertheless exhibit overwhelming correlations with activity indicators. \hl{On a related note, the traditional activity indicators we used are only imperfect proxies for activity signals in RVs. For example, on Sun-like stars, \lrhk~is not expected to correlate very strongly with activity-driven RV variations over long time scales} \citep{Milbourne19}. \hl{As such, our filtering will only be sensitive to the specific types of activity variation tracked by the indicators used.}

Following the above filtering, roughly $39\%$ of chunks remained, and were accordingly combined via an inverse variance-weighted average to produce a final set of $110$ differential RVs. While discarding $\sim61\%$ of spectra by wavelength range suggests an aggressive filtering, note for comparison that the combined width of the several thousand lines included in the DRS' G2 mask is only $\sim170$~\AA, or $\sim5\%$ of the full 1D spectrum, the latter being dominated by regions containing little Doppler information.

\subsection{Comparison of the DRS and PWGP RVs}\label{sec:DRS-vs-PWGP}

GLS periodograms of the DRS and PWGP RVs appear in the upper panel of Fig.~\ref{fig:K37_Lomb_data}, while the RVs themselves are plotted in Fig.~\ref{fig:RVs}. A quick glance at the RVs in Fig.~\ref{fig:RVs} suggests much superficial similarity; indeed, Table~\ref{tab:data-summary} confirms that the PWGP RVs have only marginally larger error bars, and a marginally smaller rms scatter, than the DRS RVs. The median and median \emph{absolute} differences between the two sets of RVs are $0.16$~\cmps\ and $1.32$~\mps, respectively. Their mutual linear (and rank) correlation coefficient is $\rho\sim0.74$ ($p\ll.001$).

The periodograms in Fig.~\ref{fig:K37_Lomb_data} reveal that both sets of RVs evince periodicities at several periods likely linked to rotationally-modulated stellar activity. However, unlike the DRS RVs, the PWGP RVs show no power at periods of $\sim100$~d or $\sim200$~d, both of which feature strongly in the periodograms of the \lrhk\ and FWHM time series, suggesting some degree of successful activity mitigation by the PWGP approach. The activity mitigation is confirmed in Table~\ref{tab:RV_correlations}, which summarises correlation between the DRS/PWGP RVs and various stellar activity indicators. Whereas the DRS RVs show both strong and significant non-parametric correlations with four different stellar activity indicators, these correlations are wholly absent from the PWGP RVs.

Intriguingly, there is far more power at Kepler-37d's period in the PWGP RVs than the DRS RVs. Even in the PWGP case, though, this power still falls below a $10\%$ FAP threshold, \hl{and is exceeded in power by nearby periodicities around $30$, $45$ and $51$~d}. If nothing else, this emphasises how weak Kepler-37d's putative signal is, and how important it will be to model the data as carefully as possible.

\section{RV modelling}\label{sec:methods2}

We now describe our modelling of the RVs extracted via the DRS or the PWGP approach. We discuss our modelling of Keplerian signals (Section~\ref{sec:modelling-kepler}); our stellar activity modelling with a GP fitted to multiple time series (Section~\ref{sec:modelling-activity}); the Bayesian framework we used for all model and parameter inference (Section~\ref{sec:modelling-Bayes}); and the practical implementation of this Bayesian inference via the \texttt{PolyChord} algorithm (Section~\ref{sec:modelling-polychord}). 

\subsection{Keplerian signals and RV offsets}\label{sec:modelling-kepler}

We model the contribution of $N_p$ orbiting planets to observed RVs via a standard prescription, ignoring planet-planet interactions (for more details, see e.g.\ \citealt{seager2011} or \citealt{perryman2011}):
\begin{equation}
    {\rm RV_{Kepler}}(t) = \sum_{p=1}^{N_p}  K_p \left[  \cos(\nu_{p}(t) + \omega_p) + e_p \cos(\omega_p)\right],
    \label{equ:keplerians}
\end{equation}
where $K_p$ is the RV semi-amplitude, $\nu_{p}$ the true anomaly, $e_p$ the orbital eccentricity, {and $\omega_p$ the argument of periastron} of the $p$\th\ planet, respectively. The true anomaly, characterising the time-dependent angle between a planet and its periapsis position, is computed via:
\begin{equation}
\tan \left(\frac{\nu_p (t) }{2}\right) = \sqrt {\frac{{1 + e_p}}{{1 - e_p}}} \tan \left( \frac{E_p(t)}{2} \right).
\end{equation}
The eccentric anomaly $E_p(t)$, in turn, is found by numerical solution of the following non-linear equation (`Kepler's equation'):
\begin{equation}
E_p(t) - e_p\sin E_p(t) = \frac{2\pi (t - {T_{0,p}})}{P_p} \equiv M_p(t),
\end{equation}
where $T_{0,p}$ defines the time at which the $p$\th\ planet is located at a particular reference point (e.g.\ periapsis), $P_p$ is that planet's orbital period, and $M_p(t)$ is defined as the mean anomaly. In practice, it is sometimes more convenient to parametrize the orbit using an angular parameter instead of using a reference point in its orbit -- e.g., $M_{0,p}$, the mean anomaly at some reference time. In total, then, there are five observables which we can fit for a single orbiting planet, on the basis of RV measurements alone: $e_p$, $P_p$, $T_{0,p}$ (or $M_{0,p}$), $\omega_p$, and $K_p$. The latter is related to the stellar and planetary masses via
\begin{equation}
K_p = {\left( {\frac{{2\pi G}}{P_p}} \right)^{1/3}}\frac{{{M_{{p}}}\sin i_p}}{{{{({M_ \star } + {M_{{p}}})}^{2/3}}}}\frac{1}{{{{(1 - {e_p^2})}^{1/2}}}},
\end{equation}
where $i_p$ denotes the angle of orbital inclination. We solve for $M_p$ from $K_p$ by making the common approximation that $M_p\ll M_\star$.

In addition to Keplerian signals, we also allowed for the possibility that each time series contained long-term trends (due e.g.\ to a long-term activity cycle, a distant orbiting body, instrumental drifts, etc.), which we modelled with polynomials up to second order. However, all our initial tests showed that (i) polynomial coefficients of first order and above were consistent within $1\sigma$ with zero in all time series, and (ii) models containing first- or second-order polynomial trends had less favourable evidence values than those containing offsets only. Therefore we restrict ourselves here to considering models with only an offset for RV, BIS and \lrhk\ time series -- hereafter denoted $\gamma_{\rm RV}$, $\gamma_{\rm BS}$, $\gamma_{\rm HK}$ -- without higher-order polynomial components.

\subsection{Stellar activity}\label{sec:modelling-activity}

In cases where we modelled stellar activity explicitly (as opposed to assuming it could be neglected, or at least accounted for via an additive white-noise `jitter' term), we used the GP framework developed by \citet{Rajpaul2015}, hereafter R15, to model RVs simultaneously with \lrhk\ and \bis\ observations -- the latter two time series being sensitive to activity-induced variability, but \emph{not} planetary signals. In short, this framework assumes that all observed stellar activity signals are generated by some underlying latent function $G(t)$ and its derivatives; this function, which is not observed directly, is modelled with a GP \citep{rasmussen2006,roberts2013}. 

Following R15, activity variability in the RV, \bis, and \lrhk\ time series can be modelled as:
\begin{align}
\Delta \rm{RV} &= V_c G(t) + V_r G'(t), \\
\bis &= B_c G(t) + B_r {G'}(t), \rm{ and} \\
\lrhk &= L_c G(t), 
\end{align}
respectively, where ${G'}(t)=dG/dt$. The coefficients $V_c$, $B_c$, $L_c$, $V_r$, and $B_r$  are free parameters relating the individual observations to the unobserved Gaussian process $G(t)$. The first three coefficients pertain to convective blueshift suppression, and the latter two to rotationally-modulated signals (hence the subscripts). $G(t)$ itself can be loosely interpreted as representing the projected area of the visible stellar disc covered in active regions at a given time \citep{aigrain2012}. The GP describing $G(t)$ is assumed to have zero mean and covariance matrix $\mathbf{K}$, where $K_{ij} = \gamma(t_i,t_j)$. As in R15, we adopt the following quasi-periodic (QP) covariance kernel function:
\begin{equation}
\label{eq:K-QP}
\gamma ({t_i},{t_j}) = \exp \left[ { - \frac{{{{\sin }^2}\left[ {\uppi ({t_i} - {t_j})/{P_{{\rm{GP}}}}} \right]}}{{2\lambda _{\rm {p}}^2}} - \frac{{{{({t_i} - {t_j})}^2}}}{{2\lambda _{\rm{e}}^2}}} \right],
\end{equation}
where $P_{\rm{GP}}$ is the period of the quasi-periodic activity signal, $\lambda_{\rm{p}}$ is the inverse harmonic complexity of the signal (such that signals become sinusoidal for large values of $\lambda_{\rm{p}}$, and show increasing complexity/harmonic content for small values of $\lambda_{\rm{p}}$), and $\lambda_{\rm{e}}$ is the time-scale over which activity signals evolve. This QP covariance kernel has been widely used to model stellar activity signals in both photometry and RVs \citep[e.g.][]{haywood2014b,Rajpaul2015,grunblatt2015,bonfils2018temperate}. Full expressions for covariances between the three different observables modelled are given in R15.

By modelling multiple activity-sensitive time series simultaneously, more information can be gleaned on activity signals in RVs,\footnote{\hl{This would not be true for a hypothetical indicator that was somehow independent of activity-driven variations in RVs; given the correlations in Table} \ref{tab:RV_correlations}, \hl{such a concern is not warranted here. We also note that it would also not be advantageous to model multiple activity indicators that contained little independent information.}} compared to approaches exploiting only simple correlations between RVs and (typically) one activity indicator \citep{gilbertson20}. In general, we would not advocate using a GP to model activity in RVs alone. Despite the convenience of such approaches \citep[e.g.][]{radvel18}, even if good prior constraints on hyper-parameters are available, there is little safeguard against fitting signals unrelated to stellar activity in order to achieve an optimal data likelihood. The term `over-fitting' might be subtly misleading in such cases, as model residuals could be perfectly consistent with the formal error bars, even if non-activity signals were inadvertently absorbed. 

As our framework uses GP draws and derivatives thereof as basis functions for modelling available time series, it avoids issues inherent in e.g.\ sinusoidal or other simple parametric models, the inappropriate use of which could easily lead to the introduction of correlated signals into model residuals. The GP basis functions could in principle take any form, although in the GP framework their properties are constrained by the data themselves, and by reasonable prior assumptions about the quasi-periodic nature of stellar activity signals. The GP framework also incorporates the so-called $FF'$ formalism directly as a special case \citep{aigrain2012}; the former approach may be thought of as a generalization of the latter. For a recent, in-depth study demonstrating advantages of R15's GP framework over a number of other approaches to modelling stellar activity, including the $FF'$ method and a multi-harmonic method, see \citet{ahrer21}.

In addition to the parameters associated directly with stellar activity in our GP model, we also included white-noise `jitter' parameters for each time series --  which we denote $\sigma_{\rm RV}^+$, $\sigma_{\rm BS}^+$, $\sigma_{\rm HK}^+$ -- and which were added in quadrature to the formal error bars for each observation. Under the GP model, these white-noise jitter parameters were intended to encapsulate activity-induced and other signals that are not adequately captured by the GP model, \hl{whether because of flaws in the assumed relationship between RVs and indicators, or because certain signals simply do not show up in chosen activity indicators} (e.g.\ stellar pulsation signals). In cases where we did \emph{not} use a GP, the jitter parameters were intended to encapsulate \emph{all} signals (including stellar activity-related ones) that could not be adequately modelled via Keplerian terms.

\subsubsection{Summary of modelling approaches used in this work}

As already noted, we consider in this work RVs extracted via two independent methods, and in each case consider the effect of using a GP to model stellar activity vs.\ using only a white noise jitter term. Moreover, for every such approach, we consider many different combinations of Keplerian terms being included in the overall model. Therefore, to aid the reader, we summarise here the four main modelling approaches we take in this work, where we use `approach' as shorthand for one particular combination of RV data set and stellar activity model, regardless of the Keplerian terms included:
\begin{enumerate}
    \item[(I)] DRS RVs -- no activity modelling;
    \item[(II)] DRS RVs -- GP activity modelling;
    \item[(III)] PWGP RVs -- no activity modelling; and
    \item[(IV)] PWGP RVs -- GP activity modelling.
\end{enumerate}

We shall often refer to these different approaches using the uppercase numerals (I)--(IV). Adopting the language used earlier in the paper, we may characterise these approaches thus: (I) no activity modelling or mitigation; (II) activity modelling only; (III) activity mitigation only; and (IV) activity mitigation combined with activity modelling. Of course, even in case (I) there is a degree of activity mitigation taking place, since e.g.\ the masks used by the DRS are designed to avoid lines known to be most susceptible to stellar activity-induced variability. Similarly, a white noise model that can (in principle) `absorb' some stellar activity variability is perhaps not ignoring activity completely. Therefore we use the terms `modelling' and `mitigation' in a largely relative sense, where it is understood that case (I) represents our baseline.

\subsection{Bayesian model and parameter inference}\label{sec:modelling-Bayes}

We use Bayesian inference to evaluate the relative posterior probabilities of (i) different models, and (ii) of different parameter values within given models. We summarise here the relevant formalism. 

\begin{table*}
    \centering
    \caption{Jeffreys' scale \citep{jeffreys61} for interpreting Bayesian evidence ratios (Bayes factors). A value $\mathcal{R}_{ij}>1$ means that model $\mathcal{M}_i$ is favoured more strongly by the data under consideration than model $\mathcal{M}_j$. We give the scale both in terms of Bayes factors and (natural) logarithms of the Bayes factors. We find the former more intuitive to interpret directly, though \texttt{PolyChord} outputs the latter, and is more convenient for evidences spanning many orders of magnitudes.}
    \label{tab:jeffreys}
    \begin{tabular}{c  c c}
    \hline
        Bayes factor & Log Bayes factor  & Strength of evidence  \\ \hline
        $\mathcal{R}_{ij}<10^0$ & $\ln\mathcal{R}_{ij}<0$ & Negative (supports $\mathcal{M}_j$) \\
        $10^0 < \mathcal{R}_{ij} < 10^{1/2}$ & $0 < \ln\mathcal{R}_{ij} < 1.15$ & Barely worth mentioning\\ 
        $10^{1/2} < \mathcal{R}_{ij} < 10^1$ & $1.15 < \ln\mathcal{R}_{ij} < 2.30$ & Substantial \\
        $10^1 < \mathcal{R}_{ij} < 10^{3/2}$ & $2.30 < \ln\mathcal{R}_{ij} < 3.45$ & Strong \\
        $10^{3/2} < \mathcal{R}_{ij} < 10^2$ & $3.45 < \ln\mathcal{R}_{ij} < 4.61$ & Very strong \\
        $\mathcal{R}_{ij}>10^2$ & $\ln \mathcal{R}_{ij}>4.61$  & Decisive
        \\ \hline
    \end{tabular}
\end{table*}

Bayes' Theorem relates the posterior probability $P(\Theta | \mathcal{D}, \mathcal{M}) \equiv \mathcal{P}(\Theta)$ of some parameters $\Theta$, given data $\mathcal{D}$ and a model $\mathcal{M}$, to
\begin{enumerate}
    \item the probability of $\Theta$, $P(\Theta | \mathcal{M}) \equiv \pi(\Theta)$, given $\mathcal{M}$;
    \item the probability of $\mathcal{D}$, $P(\mathcal{D} | \Theta, \mathcal{M})\equiv\mathcal{L}(\Theta)$, given $\Theta$, $\mathcal{M}$; and
    \item the probability of $\mathcal{D}$, $P(\mathcal{D}|\mathcal{M}) \equiv \mathcal{Z}$,  given $\mathcal{M}$.
\end{enumerate}
Bayes' theorem may be written:
\begin{equation}
    P(\Theta | \mathcal{D}, \mathcal{M}) = \frac{P(\mathcal{D} | \Theta, \mathcal{M})\ P (\Theta | \mathcal{M})}{P(\mathcal{D}|\mathcal{M})},
    \label{eq:bayes}
\end{equation}
or, using the notation from \citet{feroz2014}, simply
\begin{equation}
    \mathcal{P}(\Theta)= \frac{\mathcal{L}(\Theta)\ \pi(\Theta)}{\mathcal{Z}}.
    \label{eq:bayes2}
\end{equation}
$\mathcal{P}(\Theta)$ is the posterior probability density of the model parameters; $\mathcal{L}(\Theta)$ is the likelihood of the data, and $\pi(\Theta)$ is the parameter prior. The term in the denominator, $\mathcal{Z}$, is usually referred to as the marginal likelihood, model likelihood, or Bayesian evidence -- it represents the factor required to normalize the posterior over the entire domain of $\Theta$, i.e.\ $\mathcal{Z}=\int \mathcal{L}(\Theta)\pi(\Theta) d\Theta$. In general, $\mathcal{Z}$ is notoriously difficult to compute \citep{nelson20}. Fortunately, as this term is independent of the parameters $\Theta$, it can be ignored for parameter inference problems, where samples may be drawn from the unnormalized posterior only, as happens e.g.\ with standard MCMC methods.

For the far more challenging problem of model inference (selection), however, the marginal likelihood or evidence plays a central role.  As the evidence may be interpreted as the likelihood averaged over the prior, it is generally larger in a model where more of its total parameter space is associated with high likelihoods, and smaller for a model where large areas of parameter space have low likelihood values, even if the likelihood function is sharply peaked. Thus the evidence serves both to penalise `fine tuning' of a model against observed data, and to automatically and quantitatively implement Occam's Razor or principle of parsimony \citep[e.g.][]{occam2000,feroz2014}.

One can evaluate the relative posterior probabilities of two models $\mathcal{M}_i$ and $\mathcal{M}_j$, given data $\mathcal{D}$, by computing the ratio of their respective posterior probabilities; this ratio is also known as the Bayes factor, which we denote $\mathcal{R}_{ij}$: 
\begin{equation}
    \mathcal{R}_{ij} = \frac{P(\mathcal{M}_i | D)}{P(\mathcal{M}_j | D)} = \frac{P(\mathcal{D} | \mathcal{M}_i)\ P(\mathcal{M}_i)}{P(\mathcal{D} | \mathcal{M}_j)\ P(\mathcal{M}_j)} = \frac{\mathcal{Z}_i }{\mathcal{Z}_j } \frac{P(\mathcal{M}_i) }{P(\mathcal{M}_j)}. 
    \label{equ:post_distr_comparison}
\end{equation}
The relative prior probability of the two models, $\frac{P(\mathcal{M}_i) }{P(\mathcal{M}_j)} $, is usually set to unity (unless there happens to be some information available that would suggest favouring one model over the other \emph{a priori}), in which case $\ln \mathcal{R}_{ij} = \ln \mathcal{Z}_i - \ln \mathcal{Z}_j$. This is the approach we take throughout the analysis presented in this paper.

To decide whether the relative posterior probabilities favour one model over the other, we make use of the Jeffreys scale given in Table \ref{tab:jeffreys}. We emphasize, though, that a Bayes factor $\mathcal{R}_{ij}$ can only be used to \emph{compare} two models: a large value of $\mathcal{R}_{ij}$ may certainly lead to the model or hypothesis $\mathcal{M}_j$ being rejected, but it does not prove that $\mathcal{M}_i$ is `correct' in an absolute sense, which would in principle require evaluating \emph{all} (infinitely many) alternatives. We recall here the often-quoted words of statistician George Box \citep{box1987}: `All models are approximations. Essentially, all models are wrong, but some are useful. However, the approximate nature of the model must always be borne in mind.'

\subsubsection{Likelihood function}

In common with almost all other RV modelling efforts, we make the assumption that our model residuals are drawn from a multivariate Gaussian distribution. This may be justified via the central limit theorem, which establishes that the sum or average of many independent random variables tends towards a Gaussian distribution, even if the original variables are not normally distributed \citep{CLT-history}. Indeed, measurement errors in astronomical experiments usually contain contributions from many independent sources -- photon noise, thermo-mechanical noise, calibration errors, telescope and detector effects, atmospheric effects, etc.\ -- so this assumption is usually well-founded, strongly non-Gaussian outliers due e.g.\ to cosmic ray strikes notwithstanding. From an information theoretic perspective, the maximum entropy principle may also be used to show that in the absence of detailed knowledge of the effective noise distribution (other than assuming it has finite variance), a Gaussian distribution would be the most conservative choice, i.e.\ maximally non-committal about missing information \citep{gregory2005,sivia06}. 

Given our assumption of Gaussianity, the logarithmic likelihood of our data may be computed via the familiar expression
\begin{equation}
\ln {{\cal L}(\Theta)} = -  \tfrac{N}{2} \ln 2\upi - \tfrac{1}{2}\ln \det {{\bf{K}}} - \tfrac{1}{2}{{\bf{r}}}^{\rm{T}}{{\bf{K}}}^{ - 1}{{\bf{r}}}  ,
\label{eq:NLL_GP}
\end{equation}
where $\mathbf{K}\in\mathbb{R}^{N \times N}$ defines the covariance between all possible pairs of observations, $\mathbf{r}\in \mathbb{R}^N$ is a vector of residuals, and $N=N_{\rm RV}+N_{\rm BS}+N_{\rm HK}$ is the total number of observations of all types considered.

In cases where we used a GP to model stellar activity variability simultaneously across RV, BIS, and \lrhk measurements, $\bf{K}$ was computed via expressions given in Section~3.3 and Appendix A2 in R15, thus encoding activity-related covariances between all observations in all three time series, as well as white noise variance (observational error plus jitter parameter, added in quadrature) associated with individual observations. The vector of residuals, $\bf{r}$, was computed by subtracting the appropriate mean function from each type of observation -- a constant offset plus a varying number of Keplerian terms for RVs, and constant offsets only for BIS and \lrhk\ -- then concatenating the three sets of residuals into a single vector.

In the simpler cases where we did \emph{not} use a GP to model stellar activity, and instead assumed that all non-planetary variability in RVs could be interpreted as uncorrelated (white) noise, $\bf{K}$ was a diagonal matrix, with each diagonal element encoding white noise variance (observational error plus jitter, as before) associated with particular observations. The vector of residuals, $\bf{r}$, was computed as before. We emphasize that under this (non-GP) modelling approach, RVs, BIS, and \lrhk\ observations were all considered to be independent; the only parameters relating specifically to the BIS and \lrhk\ observations were constant offsets and white noise jitter parameters. Nevertheless, we deliberately included all three types of observations in our modelling. This had no effect on inference about RV-specific parameters, e.g.\ Keplerian parameters, but it \emph{did} have an important effect on the computation of model evidences, since $\mathcal{Z}=P(\mathcal{D}|\mathcal{M})$ always depends on the data being modelled. In other words, in order to compare the evidences of models with and without a GP stellar-activity component, it was imperative that the data in question were identical across all models. Essentially, including the BIS and \lrhk\ observations in the non-GP modelling ensured the model evidences were correctly normalised to allow comparison with the GP-based models. Thus, we could answer questions such as: `to what extent do the data support using a more complex GP model to model variability in three (possibly related) time series?'

\subsubsection{Parameter priors}\label{sec:param-prior}

We use the symbols $\mathcal{N}$, $\mathcal{U}$, and $\beta$ to denote normal, uniform and Beta distributions, respectively, in each case with standard parametrizations. We denote a Jeffreys prior over some parameter $0<a \le \theta\le b$ via $\mathcal{J}(a,b)$, such that
\begin{equation}
    p(\theta) = \frac{1}{\theta} \times \frac{1}{\ln\left(a/b\right)}.
\end{equation}
This prior (also known as a log uniform prior) is scale invariant, so is an appropriate choice for parameters whose scale is not known \emph{a priori}; by contrast, a standard uniform prior is inherently biased to larger parameter values \citep{gregory2005}. The Jeffreys prior is not suitable, however, for parameters whose minimum allowable value is zero, in which case the Jeffreys prior is not normalisable. In such cases, a modified Jeffreys prior over some parameter $0\le\theta\le \theta_{\rm max}$, which we denote $\rm{mod} \mathcal{J}(\theta_0,\theta_{\rm max})$, is more suitable:
\begin{equation}
    p(\theta) = \frac{1}{\theta+\theta_{\rm max}} \times \frac{1}{\ln\left(1+\theta_{\rm max}/\theta_0\right)};
\end{equation}
this mirrors the Jeffreys prior closely for large values of $\theta$, though resembles a uniform prior when $\theta<\theta_0$. 

The priors we placed on the parameters of the three planets transiting \mystar, the putative TTV planet (Kepler-37e), and any other possible planets in our models are given in Table~\ref{tab:planet-priors}. Where informative prior constraints on a (real or putative) planet's expected RV semi-amplitude were not available, we set the maximum allowable RV semi-amplitude to be the standard deviation of the observed RVs. This was essentially a computational convenience, albeit an easily justifiable one: given the known presence of white noise, correlated stellar and other nuisance signals, multiple known planets, etc., it is not plausible that a single planet could account for the entirety of the observed RV variability. We also confirmed, in preliminary runs, that posterior distributions for $K_i$ never peaked near the prior upper limit. We allowed the maximum periods of putative planets to be slightly more than double the temporal baseline of our HARPS-N observations.

Apart from the priors we placed on individual planetary parameters, we also wished to impose the prior restriction that planetary orbits should not be unstable. Accordingly, we checked the mutual orbital stability of all pairs of Keplerian orbits, using a criterion introduced by \citet{stabilityGladman}, and often used in RV modelling \citep[e.g.][]{Malavolta2017}: $\Delta > 2 \sqrt{3}\ R_H(i,j)$ where $\Delta = a_j - a_i$ is the difference between the semi-major axis of the $i$\th\ and $j$\th\ planet, and $R_H(i,j)$ the planets' mutual Hill radius. Any combination of planet parameters giving rise to at least one unstable orbit was rejected by setting the associated data likelihood  to zero.

The priors we placed on all other (i.e.\ non-planetary) parameters in our models appear in Table~\ref{tab:other-priors}; these were all, essentially, uninformative priors. For example, we only constrained the additive white noise jitter parameters to be less than the standard deviation of the time series to which they related (RV, BIS or \lrhk), for reasons analogous to our prior constraints on the RV semi-amplitudes $K_e$, $K_f$, etc. We constrained our GP hyper-parameters to a broad range likely to cover all physically-plausible possibilities. However, reasoning that we were using the GP to model quasi-periodic stellar activity signals, we did reject parameter combinations that would have resulted in a kernel that was not even quasi-periodic. In particular, we required: 
\begin{equation}
\label{eq:QP-criterion}
    \lambda_{\rm e}^2 > 3P_{\rm GP}^2 \lambda_{\rm p}^2/2\upi,
\end{equation}
which must be satisfied for the QP covariance function to have at least one non-trivial turning point \citep{vinesh-thesis}.\footnote{This constraint ultimately turned out to be superfluous, as the hyper-parameter posterior probability mass was strongly concentrated far away from regions where this constraint may have been violated (see Table~\ref{tab:param-posteriors}) However, we include the constraint here for completeness' sake.}

\begin{table}
	\centering
	\caption{Priors placed over the parameters in the Keplerian components of our RV models. As discussed in Section~\ref{sec:param-prior}, we also rejected parameter combinations giving rise to at least one pair of unstable Keplerian orbits. Additional notes: (1) we use the shorthand $s_{\rm RV}$ to denote the standard deviation of the RV time series, as given in Table~\ref{tab:data-summary}; (2) we also required periods to be sorted, i.e.\ $P_f<P_g<\ldots$; (3) see \citet{kipping13} for more details.}
	\label{tab:planet-priors}
\begin{tabular}{ccc}
\hline
Parameter & Prior & Notes \\  \hline
$K_b$ (\mps)        & $\rm{mod}\mathcal{J}(10^{-4},10^{-2})$     & Cf.\ Section~\ref{sec:planet-detectability}      \\
$K_c$ (\mps)        & $\rm{mod}\mathcal{J}(10^{-4},0.2)$     & Cf.\ Section~\ref{sec:planet-detectability}      \\
$K_i$ (\mps)        & $\rm{mod}\mathcal{J}\left(10^{-1}, s_{\rm RV}\right)$   &  $i\in\{d,e,f,\ldots\}$; (1)      \\ \hline
$P_b$ (d)        &  $13.367020$    & Fixed; cf.~Table~\ref{tab:planet-properties}      \\
$P_c$ (d)        & 21.301848     & Fixed; cf.~Table~\ref{tab:planet-properties}      \\
$P_d$ (d)        & 39.7922622     & Fixed; cf.~Table~\ref{tab:planet-properties}      \\
$P_e$ (d)        & $\mathcal{N}(51.196,1)$     &  cf.~Table~\ref{tab:planet-properties}     \\
$P_i$ (d)       & $\mathcal{J}(0.5,4000)$     & $i\in\{f,g,\ldots\}$; (2)       \\ \hline
$e_i$        & $\beta(0.867, 3.03)$     & $i\in\{b,c,\ldots\}$; (3)        \\ \hline
$T_{0,b}$ (BJD) & 2455017.0473 & Fixed; cf.~Table~\ref{tab:planet-properties}\\
$T_{0,c}$ (BJD) & 2455024.83997 & Fixed; cf.~Table~\ref{tab:planet-properties}\\
$T_{0,d}$ (BJD) & 2455008.24982 & Fixed; cf.~Table~\ref{tab:planet-properties}\\
$M_i$ (rad)        & $\mathcal{U}(0, 2\upi)$     & $i\in\{e,f,\ldots\}$        \\ \hline
$\omega_i$ (rad)     & $\mathcal{U}(0,2\upi)$ & $i\in\{b,c,\ldots\}$ \\ \hline
\end{tabular}
\end{table}

\begin{table}
	\centering
	\caption{Priors placed over the non-planetary -- white noise jitter, offsets, and GP i.e.\ stellar activity-related -- components of our models. Here, we use the shorthand $\mu_i$ and $s_i$ to denote the mean and standard deviation of a particular set of measurements, e.g.\ $\mu_{\rm HK}$ is the mean value of the \lrhk time series, and $s_{\rm BS}$ is the standard deviation of the BIS time series.}
	\label{tab:other-priors}
\begin{tabular}{ccc}
\hline
Parameter & Prior & Notes \\  \hline
$\sigma_{\rm RV}^+$ (\mps)    & $\rm{mod}\mathcal{J}(10^{-1},s_{\rm RV})$     &   Cf.\ Table~\ref{tab:data-summary}    \\
$\sigma_{\rm BIS}^+$ (\mps)       & $\rm{mod}\mathcal{J}(10^{-1},s_{\rm BS})$     & Cf.\ Table~\ref{tab:data-summary}       \\
$\sigma_{\rm HK}^+$             & $\rm{mod}\mathcal{J}(10^{-3},s_{\rm HK})$     & Cf.\ Table~\ref{tab:data-summary}       \\ \hline
$\gamma_{\rm RV}$ (\mps)        & $\mathcal{N}\left(\mu_{\rm RV},s_{\rm RV} \right)$     & \bl       \\ 
$\gamma_{\rm BS}$ (\mps)        &  $\mathcal{N}\left(\mu_{\rm BS},s_{\rm BS} \right)$     & \bl      \\ 
$\gamma_{\rm HK}$        &  $\mathcal{N}\left(\mu_{\rm HK},s_{\rm HK} \right)$     &  \bl     \\ \hline
$P_{\rm GP}$ (d)        & $\mathcal{J}(10,100)$     & See also equation~\ref{eq:QP-criterion}      \\
$\lambda_{ e}$ (d)        & $\mathcal{J}(10,1000)$     & See also equation~\ref{eq:QP-criterion}      \\
$\lambda_{ p}$ (\mps)        & $\mathcal{J}(10^{-1},10)$     & See also equation~\ref{eq:QP-criterion}      \\ \hline
$V_c$ (\mps)        & $\mathcal{N}\left(0,\sigma_{\rm RV} \right)$     &    \bl   \\
$V_r$ (\mps)        & $\mathcal{N}\left(0,\sigma_{\rm RV} \right)$     &    \bl   \\
$B_c$ (\mps)        & $\mathcal{N}\left(0,\sigma_{\rm BS} \right)$     &    \bl   \\
$B_r$ (\mps)        & $\mathcal{N}\left(0,\sigma_{\rm BS} \right)$     & \bl      \\
$L_c$        & $\mathcal{N}\left(0,\sigma_{\rm HK} \right)$     &  \bl     \\
\hline
\end{tabular}
\end{table}

\subsection{\pchord}\label{sec:modelling-polychord}

Throughout this study, we use \pchord\ for all parameter and model inference. \pchord\ is a state-of-the-art nested sampling algorithm, designed to work well even with parameter spaces with very large dimensionality \citep{polychord15}. A short discussion illuminating its favourable properties compared with \texttt{MultiNest} -- its direct predecessor, and a nested sampling tool that has been widely used in exoplanet studies \citep{multinest2008, multinest2009, feroz2014} -- can be found in \citet{hall2018}. A detailed study testing \pchord\ and then leveraging it specifically for joint modelling of exoplanets and stellar activity in RVs (as in the present study) is given by \citet{ahrer21}. 

\pchord\ is written in C++ and Fortran, though we called it via the \texttt{pypolychord} Python wrapper. We generally left \pchord's sampling parameters at their default values, except for the stopping (precision) criterion, which we changed from the default $10^{-3}$ to a \emph{much} more stringent $10^{-12}$: we found the default value led to evidence values that often differed significantly from run to run \citep{ahrer21}. As \pchord\ natively supports MPI (Message Passing Interface) parallelisation, we were able to run it on a high-performance computing platform, typically using several hundred CPU cores simultaneously. This made it feasible for us to compute accurate Bayesian evidences even for models with high dimensionality and computationally-intensive GP components to evaluate.% a single run of \pchord\ in such cases often required $\sim\mathcal{O}(10^8)$ likelihood evaluations before convergence was achieved.

To obtain robust estimates of the uncertainty in computed model evidences, we considered (i) \pchord's internal estimates provided by individual runs, and (ii) the standard error in the mean of model evidences returned from multiple \pchord\ runs, and adopted the larger of the two as our most plausible uncertainty estimate \citep[cf.][]{nelson20,ahrer21}.

\section{Results and discussion}\label{sec:results}

\begin{table*}
\centering
\caption{Relative Bayesian evidences for models including different numbers of Keplerian terms. For ease of interpretation, the evidence for the GP model including Kepler-37d as the only Keplerian term is here defined to be zero for both the DRS RVs ($\mathcal{Z}_*$) and the PWGP RVs ($\mathcal{Z}_\dagger$); \emph{all} other evidences are given relative to one of these values. Evidences for models for different data (DRS vs PWGP RVs) can not be compared meaningfully. Uncertainties correspond to the standard error on the mean evidence across five \textsc{PolyChord} runs, or to the highest individual run error provided by \textsc{PolyChord} (the greater of the two).}
\label{tab:evidences}
\begin{tabular}{c|cccc|cccc}
\hline
\multirow{3}{*}{Keplerians} & \multicolumn{2}{c}{(I) DRS; no} & \multicolumn{2}{c|}{(II) DRS + GP} & \multicolumn{2}{c}{(III) PWGP; no} & \multicolumn{2}{c}{(IV) PWGP + GP} \\

 & \multicolumn{2}{c}{activity model} & \multicolumn{2}{c|}{activity model} & \multicolumn{2}{c}{activity model} & \multicolumn{2}{c}{activity model} \\

                  & $\ln(\mathcal{Z}/\mathcal{Z}_*)$                  & $\sigma(\ln\mathcal{Z})$             & $\ln(\mathcal{Z}/\mathcal{Z}_*)$                & $\sigma(\ln\mathcal{Z})$            & $\ln(\mathcal{Z}/\mathcal{Z}_\dagger)$                   & $\sigma(\ln\mathcal{Z})$               & $\ln(\mathcal{Z}/\mathcal{Z}_\dagger)$                & $\sigma(\ln\mathcal{Z})$             \\ \hline

\bl & $-100.524$           & 0.065           & $-2.834$           & 0.176          & $-92.371$             & 0.063             & $-6.138$           & 0.128           \\

$b$                  & $-101.775$           & 0.078           & $-4.178$           & 0.160          & $-94.891$             & 0.082             & $-6.010$           & 0.156           \\

$c$                  & $-102.041$ & 0.077           & $-4.030$           & 0.162          & $-94.166$             & 0.082            & $-6.237$           & 0.142           \\

$d$                  & $-101.377$           & 0.086           & $\equiv0$            & 0.131          & $-88.865$             & 0.086             & $\equiv0$            & 0.124           \\

$e$                  & $-100.202$           & 0.095           & $-1.646$           & 0.157          & $-90.501$             & 0.093             & $-3.689$           & 0.154           \\

$f$                  & $-97.563$            & 0.127           & $-3.159$           & 0.238          & $-89.786$             & 0.141             & $-3.127$           & 0.167           \\

$b,d$                & $-102.248$           & 0.142           & $-1.829$           & 0.156          & $-90.669$             & 0.145             & $-0.362$           & 0.142           \\

$c,d$                & $-102.556$           & 0.142           & $-1.773$           & 0.214          & $-90.823$             & 0.145             & $-0.465$           & 0.152           \\

$d,e$                & $-100.939$           & 0.085           & $-0.507$           & 0.121          & $-88.394$             & 0.089             & $+0.008$            & 0.116           \\

$d,f$                & $-98.671$            & 0.194           & $-0.673$          & 0.249          & $-89.116$             & 0.101             & $-0.914$           & 0.173           \\

$d,e,f$              & $-98.239$            & 0.108           & $-0.770$           & 0.253          & $-89.351$             & 0.094             & $-1.179$           & 0.188           \\

$d,f,g$              & $-97.683$            & 0.176           & $-0.913$           & 0.284          & $-89.729$             & 0.103             & $-1.409$           & 0.228           \\ \hline
\end{tabular}
\end{table*}

\begin{table}
\centering
\caption{Bayes factors $\mathcal{R}_{d0}\equiv\mathcal{Z}_d/\mathcal{Z}_0$, quantifying the degree to which a model containing Kepler-37d was favoured vs.\ a planet-free model, under modelling approaches (I)--(IV). While these Bayes factors and uncertainties may be derived from Table~\ref{tab:evidences}, we present them here for ease of reference.}
\label{tab:evidences2}
\begin{tabular}{ccc}
\hline
Modelling approach & Bayes factor $\mathcal{R}_{d0}$ & K-37d detection\\
\hline
(I) DRS; no activity model & $0.426_{-0.043}^{+0.048}$ & \bl \\
(II) DRS + GP activity model & $17.0^{+4.2}_{-3.3}$ &  Strong\\
(III) PWGP; no activity model & $33.3^{+3.6}_{-3.3}$ & Very strong \\
(IV) PWGP + GP activity model & $463^{+90}_{-75}$ & Decisive \\
\hline
\end{tabular}
\end{table}

\begin{table}
\centering
\caption{Residual rms scatter in RV, BIS and \lrhk\ time series, under modelling approaches (I)--(IV), in each case for the model containing Kepler-37d as the only Keplerian. For comparison, the rms of the mean-subtracted DRS and PWGP RVs was $2.68$ and $2.50$~\mps, respectively (see Table~\ref{tab:data-summary}). %The rms scatters of the BIS and \lrhk\ time series are identical to those of the residuals under approaches (I) and (III). 
}
\label{tab:residuals}
\begin{tabular}{ccccc}
\hline
\multirow{3}{*}{Modelling approach} & \multicolumn{3}{c}{Residual rms} \\
 & RV    & BIS   & \lrhk &  \\
                                    & (\mps) & (\mps) &     \bl  &  \\
\hline
(I) DRS; no activity model                & 2.66     & 3.47      & 0.0227 &  \\
(II) DRS + GP activity model              & 2.08     & 2.74      & 0.0072 &  \\
(III) PWGP; no activity model             & 2.35     & 3.47      & 0.0227 &  \\
(IV) PWGP + GP activity model              & 2.24     & 2.79      & 0.0067 & \\
\hline
\end{tabular}
\end{table}

\subsection{Model selection}

Table~\ref{tab:evidences} gives the Bayesian evidences we computed for models featuring various numbers and combinations of Keplerian terms -- including ones for all three transiting planets, and the putative planet Kepler-37e -- under each of our four modelling approaches. These results are based on multiple, repeated \texttt{PolyChord} runs for each model. As this table encapsulates several of our main results, albeit in a very condensed format, we unpack a few of the main findings below. Note that our focus for now is primarily on model selection; in Section~\ref{sec:param-posterior} we shall turn our attention to more fine-grained questions about parameter values inferred under various models.

\subsubsection{Non-detection of Kepler-37b and Kepler-37c}

Regardless of the dataset or stellar activity model, models that included either only Kepler-37b or c were \emph{never} favoured over the simpler null (planet-free) models, or models additionally containing Kepler-37d. This was unsurprising, though reassuring, as we did not expect to be able to detect either of the inner two transiting planets.

\subsubsection{Non-detection of new planets}

Under modelling approach (I), representing the case of no activity mitigation or modelling, models with more `free' Keplerians (in principle, representing hitherto-unknown planets) were favoured over models without these components. However, parameter posteriors revealed that these Keplerians were almost certainly being used to absorb stellar signals. Their periods often corresponded to periodicities in the power spectra of activity indicators; they tended to have large ($e>0.2$) eccentricities; and their other orbital parameters were weakly constrained. Different runs of identical models led to similar evidence values, but with different periods sometimes being favoured. In short, following the reasoning set out in \citet{ahrer21}, we concluded that these results were indicative of inadequate stellar activity modelling, \emph{not} of the detection of genuine planets.

Under modelling approaches (II)--(IV), however, models including free Keplerians were never favoured over the model containing Kepler-37d only. In some cases, a model with one free Keplerian was at least favoured over the planet-free model. However, it always turned out that this Keplerian term was used to account for variability at $\sim100$~d or $\sim200$~d (periodicities which we have already noted likely have a stellar origin), with other orbital parameters poorly constrained.

%, and moreover as the absolute evidences for \emph{all} models under approach (III) were many orders of magnitude lower than even the worst models under approach (IV), we could very safely conclude that we did not detect any new planets.

\subsubsection{Decisive detection of Kepler-37d}

Our most interesting and striking results concern Kepler-37d. Moving from approaches (I)--(IV), the strength of the evidence supporting a model containing Kepler-37d only, versus the null model, increases significantly. In approach (I), the evidence is `barely worth mentioning,' while in the other three approaches, the evidence is `strong,' `very strong,' and `decisive,' respectively, following Jeffreys' scheme in Table~\ref{tab:jeffreys}. Whether considering the DRS RVs or the PWGP RVs, a model containing Kepler-37d as the only Keplerian contribution to RVs is decisively favoured over all alternatives. These findings are summarised quantitatively in Table~\ref{tab:evidences2} which, unlike Table~\ref{tab:evidences}, focuses on the strength of detection for Kepler-37d only, rather than numerous alternative models containing various other Keplerians.

As one might have hoped, then, mitigating and modelling stellar activity each enabled the detection of a planetary signal that would otherwise have remained `buried' under stellar nuisance signals. Moreover, combining these two approaches led to a stronger, decisive detection of said planetary signal.

\subsubsection{Non-detection of Kepler-37e}

Models containing only Kepler-37e -- or more precisely, a Keplerian with $\sim51$~d period -- were rejected compared to the null model under approaches (I), (II), and (IV); similarly, models containing both Kepler-37d and Kepler-37e were \emph{not} favoured over simpler models containing only Kepler-37d. The only tentative evidence for a Kepler-37e detection came under approach (III), where a model with Kepler-37e only (and $K_e\sim1.0\pm0.4$~\mps) was favoured over the null model. However, models including both Kepler-37d and Kepler-37e (in which case $K_e\sim0.8\pm0.4$~\mps) were not favoured over the simpler Kepler-37d-only model, with the latter being strongly favoured over the null model. Given also the overwhelmingly lower evidence for all models under approach (III) compared to approach (IV), we conclude that we did not detect Kepler-37e in our RVs.

\subsubsection{Model residual analysis}

\begin{figure}

	\includegraphics[width=\linewidth]{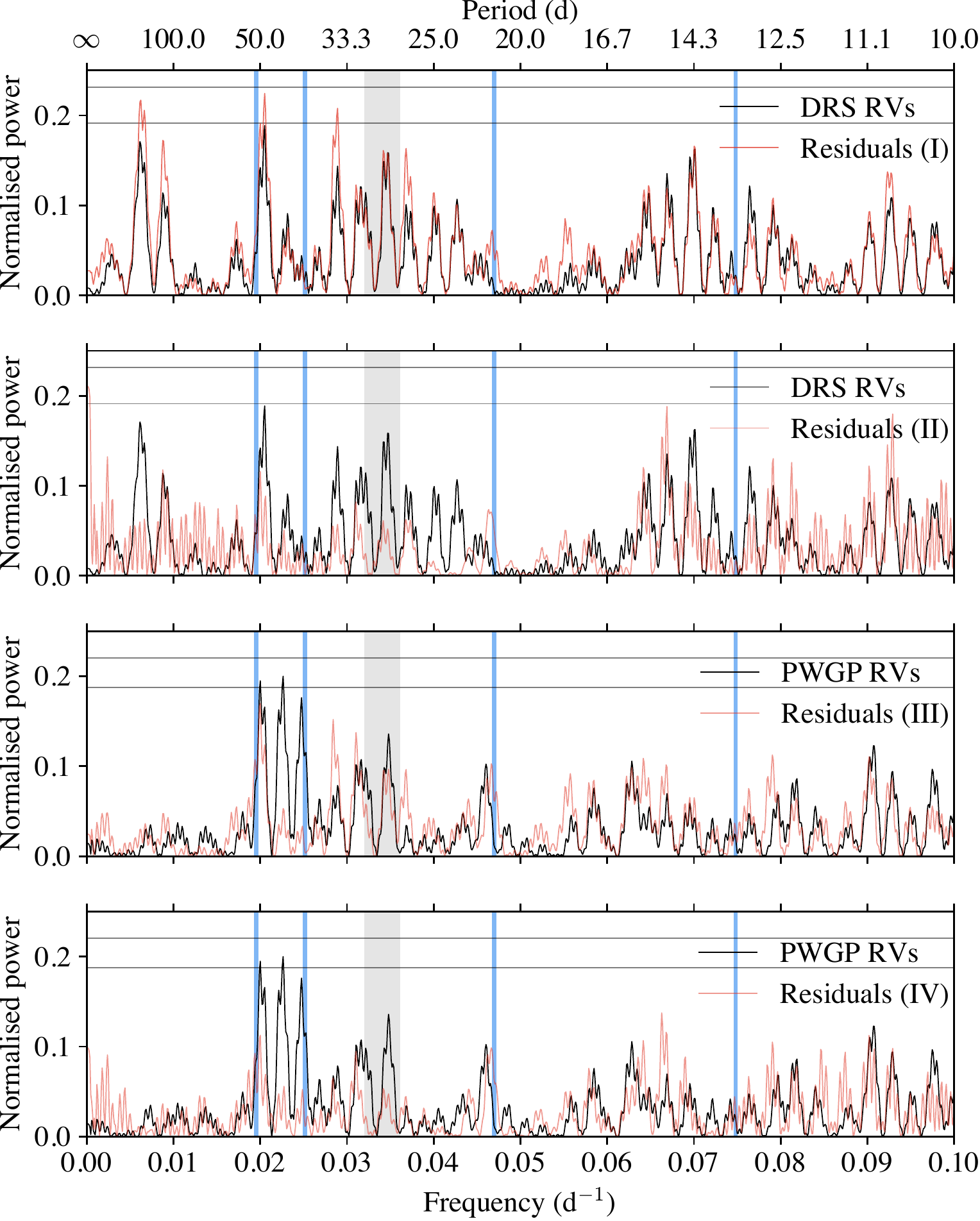}
    \caption{GLS periodogram of Kepler-37 RV residuals, after subtracting our best-fitting Kepler-37d model -- plus GP activity model, in cases (II) and (IV) -- for the DRS RVs and PWGP RVs. As in Fig.~\ref{fig:K37_Lomb_data}, vertical blue lines indicate the orbital periods of Kepler-37b through to Kepler-37e; the grey shaded box covers a $\pm2\sigma$ credible interval around Kepler-37's estimated rotation period, but now based on our GP modelling (cf.\ Table~\ref{tab:param-posteriors}). The lower and upper horizontal lines indicate estimated $10\%$ and $1\%$ FAP thresholds, respectively.}
    \label{fig:K37_Lomb}
    
\end{figure}

In Fig.~\ref{fig:K37_Lomb} we show the Lomb-Scargle periodograms of the residuals from our best-fitting, Kepler-37d-only model in approaches (I)--(IV); note that, in cases (II)--(IV), the associated model was also favoured over all others considered under the respective approach. 

Fig.~\ref{fig:K37_Lomb} shows that the residuals from approach (I) are the only ones to contain any significant (FAP $<10\%$) periodicities. Strikingly, the most significant periodicity occurs around $\sim51$~d, i.e.\ the supposed period of Kepler-37e \hl{(corresponding to the leftmost blue line in the plot; the associated frequency is $0.02$~d$^{-1}$)}. In the other three approaches, each of which entails stellar activity modelling and/or mitigation, the $51$~d periodicity is absent from the residuals. \hl{As $50.9$~d is the primary annual alias of $30$~d, we suspect the apparent $51$~d periodicity was suppressed in tandem with the $30$~d activity signal. The presence of the apparent $\sim51$~d periodicity not only in Kepler-37 RVs and photometry, but also in the \lrhk~and FWHM time series (Fig}.~\ref{fig:K37_Lomb_data}), \hl{provides further support for this periodicity/alias being activity-related. A planetary origin seems implausible, given that the periodicity is absent even from the residuals in approach (III), where one could not contend that a GP model may have inadvertently removed a Keplerian signal with this period, since no GP modelling took place.}

The absence of significant periodicities in the residuals from approaches (II)--(IV) does not prove that there are no planetary signals buried in the data. (Indeed, we know Kepler-37b and c are present, and Kepler-37d's signal has FAP~$>10\%$ even in the PWGP \emph{RVs}). On the other hand, this does at least conform with the conclusions from our more robust Bayesian model comparisons, \emph{viz}.\ that the data do not support the detection of additional Keplerian signals. \footnote{In general, we would \emph{not} advocate using periodograms as the basis for conclusions about planet detections or non-detections. We use periodograms here for a quick, intuitive way of visualising the end result of fitting models to (quasi-)periodic data, and of diagnosing glaring shortcomings in such fits.}

Additionally, autocorrelation functions of the residuals from approaches (I)--(IV) showed no significantly non-zero autocorrelations for any lags.

\subsubsection{Quantitative justification for GP modelling}

In both the DRS and PWGP cases, the Bayesian evidence for the best model without a GP was many (tens of) orders of magnitude lower than the worst model including a GP. The interpretation is that the added complexity and larger overall volume of parameter space introduced by the GP is more than compensated for by the dramatic improvements in the extent to which the GP model can explain non-Keplerian correlated variability in the data (see Table~\ref{tab:residuals}). This is despite our fairly stringent requirement that a single GP and its derivative must account for activity-related variability in RVs, BIS and \lrhk\ time series \emph{simultaneously} (see Section~\ref{sec:modelling-activity}). Moreover, the volume of parameter space associated with reasonably good fits seems to be much smaller under the non-GP models (cf.\ e.g.\ the narrow error bars on white-noise jitter parameters in Table~\ref{tab:param-posteriors}), which is indicative of inappropriate model complexity, and would contribute to the very unfavourable evidence values \citep{occam2000}.

While one could make several heuristic arguments in favour of the GP modelling -- including the need to account for the manifest though hard-to-parametrize activity contamination noted in Section~\ref{sec:DRS-vs-PWGP}, and the fact that in practice, the GP modelling enabled the most secure RV detection of Kepler-37d -- it is nevertheless interesting to be able to justify the GP's use via Bayesian model comparison.

\begin{table*}
\centering
\caption{Summary of marginalised, 1D posteriors for all parameters in modelling approaches (I)--(IV), in each case where Kepler-37d was the only Keplerian component in the model. For each parameter, the posterior median and $\pm\sigma$ credible interval around the median are given; parameters are separated into categories following the scheme in Tables~\ref{tab:planet-priors} and \ref{tab:other-priors}. The angular parameter $\omega_d$ (argument of periastron) has been transformed to the domain $[\upi,3\upi]$ to suppress boundary effects associated with the original domain $[0,2\upi]$; the posterior peaks close to the edges of the latter domain, leading to apparent though spurious bimodality.}
\label{tab:param-posteriors}
\begin{tabular} { l c c  c c c}

% 'K37_c420_d_a'; 'K37_402_d_a'; 'K37_c305_d_a'; 'K37_412_d_a'

\hline

 & (I) & (II) & (III) & (IV) \\
 Parameter & DRS; no & DRS + GP &  PWGP; no & PWGP + GP & Notes \\
  & activity model & activity model & activity model & activity model\\
\hline
{$K_d$ (\mps)}  & $0.340^{+0.039}_{-0.18}$ &     $0.86\pm 0.30$ &  $0.98^{+0.11}_{-0.18}$ & $1.22\pm 0.31   $ & $1\sigma$ agreement: (II)--(IV) only \\

{$e_d    $} & $0.126^{+0.016}_{-0.050}$ &   $0.1163^{+0.0079}_{-0.053}$ & $0.148^{+0.028}_{-0.044}$ & $0.142^{+0.026}_{-0.044}   $ &  $1\sigma$ agreement for all \\

{$\omega_d   $} & $6.4^{+1.1}_{-1.0}$ &  $6.44^{+1.8}_{-0.27}$ & $6.37^{+0.27}_{-0.62}$ & $6.39^{+0.35}_{-0.64}   $ & $1\sigma$ agreement for all\\ \hline

{$\sigma^{+}_{\rm RV}$} (\mps) & $2.344^{+0.080}_{-0.00016}$ & $1.989^{+0.078}_{-0.11}$ & $2.030^{+0.090}_{-0.035}$ & $1.96^{+0.10}_{-0.056}   $ &$1\sigma$ agreement: (II)--(IV) only\\

{$\sigma^{+}_{\rm BS}$} (\mps) & $2.63^{+0.11}_{-0.017}$ & $1.97\pm 0.31$  & $2.60^{+0.12}_{-0.0091}$ & $1.99\pm 0.30 $ & $1\sigma$ agreement: (I) \& (III), (II) \& (IV)\\

{$\sigma^{+}_{\rm HK}$} & $0.02105^{+0.00070}_{-0.00010}$ & $0.0038\pm 0.0010$ & $0.02100^{+0.00077}_{-0.00015}$ & $0.00363^{+0.00052}_{-0.00043}$  & $1\sigma$ agreement: (I) \& (III), (II) \& (IV)\\ \hline

{$\gamma_{\rm RV} (\mps)    $} & $-30651.55\pm 0.16$ & $-30651.50\pm 0.19$ & $-0.14\pm 0.15$ & $-0.11\pm 0.17$ & Cf.\ comments in Section~\ref{sec:HN-obs}\\

{$\gamma_{\rm BS} (\mps)    $} & $4.60\pm 0.20$ & $4.48\pm 0.23$ & $4.59\pm 0.21$ & $4.47\pm 0.23$ & $1\sigma$ agreement for all \\

{$\gamma_{\rm HK}  $} & $-4.8705\pm 0.0012$  & $-4.8703\pm 0.0019$ & $-4.8706\pm 0.0014$ & $-4.8702\pm 0.0020$ & $1\sigma$ agreement for all \\ \hline

{$P_{\rm GP}     $} (d) & \bl & $29.32^{+0.54}_{-0.80}$ & \bl & $29.46^{+0.57}_{-0.93}$ & $1\sigma$ agreement\\

{$\lambda_{\rm e}   $} (d) & \bl & $29.0^{+1.1}_{-4.3}$ & \bl & $29.3^{+1.2}_{-4.3}$  & $1\sigma$ agreement\\

{$\lambda_{\rm p}   $} & \bl & $0.687^{+0.027}_{-0.087}$ & \bl & $0.722^{+0.032}_{-0.098}$  & $1\sigma$ agreement\\ \hline

{$V_r   $} (\mps) & \bl & $3.49^{+1.8}_{-0.059}$ & \bl & $1.340^{+0.053}_{-0.88}$ & (II) and (IV) discrepant; cf.\ Section~\ref{sec:param-posterior} \\

{$V_c   $} (\mps) & \bl & $-1.65^{+0.20}_{-0.12}$ & \bl & $-0.78^{+0.16}_{-0.14}$  & (II) and (IV) discrepant; cf.\ Section~\ref{sec:param-posterior}  \\

{$B_r   $} (\mps) & \bl & $-5.8\pm 2.0$ & \bl & $-5.9\pm 2.2$ & $1\sigma$ agreement\\

{$B_c   $} (\mps) & \bl & $-1.76^{+0.28}_{-0.12}$ & \bl & $-1.88^{+0.28}_{-0.17}$ & $1\sigma$ agreement \\

{$L_c   $} & \bl & $-0.0253^{+0.0025}_{-0.00086}$ & \bl & $-0.0271^{+0.0030}_{-0.00081}$ & $1\sigma$ agreement\\

\hline
\end{tabular}
\end{table*}

\subsubsection{Computational burdens of our analysis}\label{sec:burden}

The most complex models we evaluated, featuring three Keplerian terms and a GP stellar activity model, required $\mathcal{O}(10^8)$ likelihood evaluations before \texttt{PolyChord} converged; the simplest models typically required $\mathcal{O}(10^6)$ likelihood evaluations. Since we considered models featuring many different combinations of Keplerian terms (including but not limited to the 12 combinations given in Table~\ref{tab:evidences}), across four different approaches, and evaluated every such combination five times, we ultimately evaluated $\mathcal{O}(10^{10})$ likelihood functions. This required tens of thousands of CPU hours on a high-performance computing platform. One might well ask whether such computationally-expensive analyses are worth the trouble. We would respond with a resounding `yes.' 

The lion's share of our computational budget was associated with inverting large covariance matrices, as required for GP likelihood evaluations. The overwhelming benefit of using a GP to model stellar activity across multiple time series is borne out by Tables~\ref{tab:evidences}~and~\ref{tab:evidences2}: if one had neglected to model stellar activity on this (nominally) inactive star, one would simply not have detected Kepler-37d. It is beyond the scope of this study to consider activity models less complex or sophisticated than our GP framework; such an in-depth investigation was, however, carried out by \citet{ahrer21}, who ultimately argued in favour of using a GP to model stellar activity across multiple time series, despite the computational burdens.

Even without the burdens of GP evaluation, computing Bayesian evidences is notoriously difficult. Yet as an approach to evaluating statistically whether one has detected a planet, it is virtually unassailable in its rigour. Compare, for example, the GLS periodograms in Fig.~\ref{fig:K37_Lomb_data}, which would by themselves have lent little or no credibility to any claim of a Kepler-37d detection. Similarly, some of our models had posterior semi-amplitudes for certain Keplerians -- e.g.\ Kepler-37e -- that were inconsistent with zero at $>2\sigma$ levels, superficially suggesting a detection, even though the model itself was ultimately rejected. A subtle caveat, though, is that Bayesian model selection can never decide whether a model is `correct' in an absolute sense -- hence the need to evaluate many competing, plausible models. 

Finally, when computing Bayesian evidences with \texttt{PolyChord}, we deliberately used very stringent convergence criteria, to try to ensure robust and consistent runs; as shown by \citet{nelson20}, \citet{ahrer21} and others, obtaining very large scatters in model evidences is often a pernicious problem when evaluating multi-Keplerian models, and could easily result in both false-positive or false-negative detections. Reassuringly, we obtained very small scatters in model evidences computed across multiple, independent runs.

% Figure: K37-d posterior
\begin{figure*}

	\includegraphics[scale=0.6]{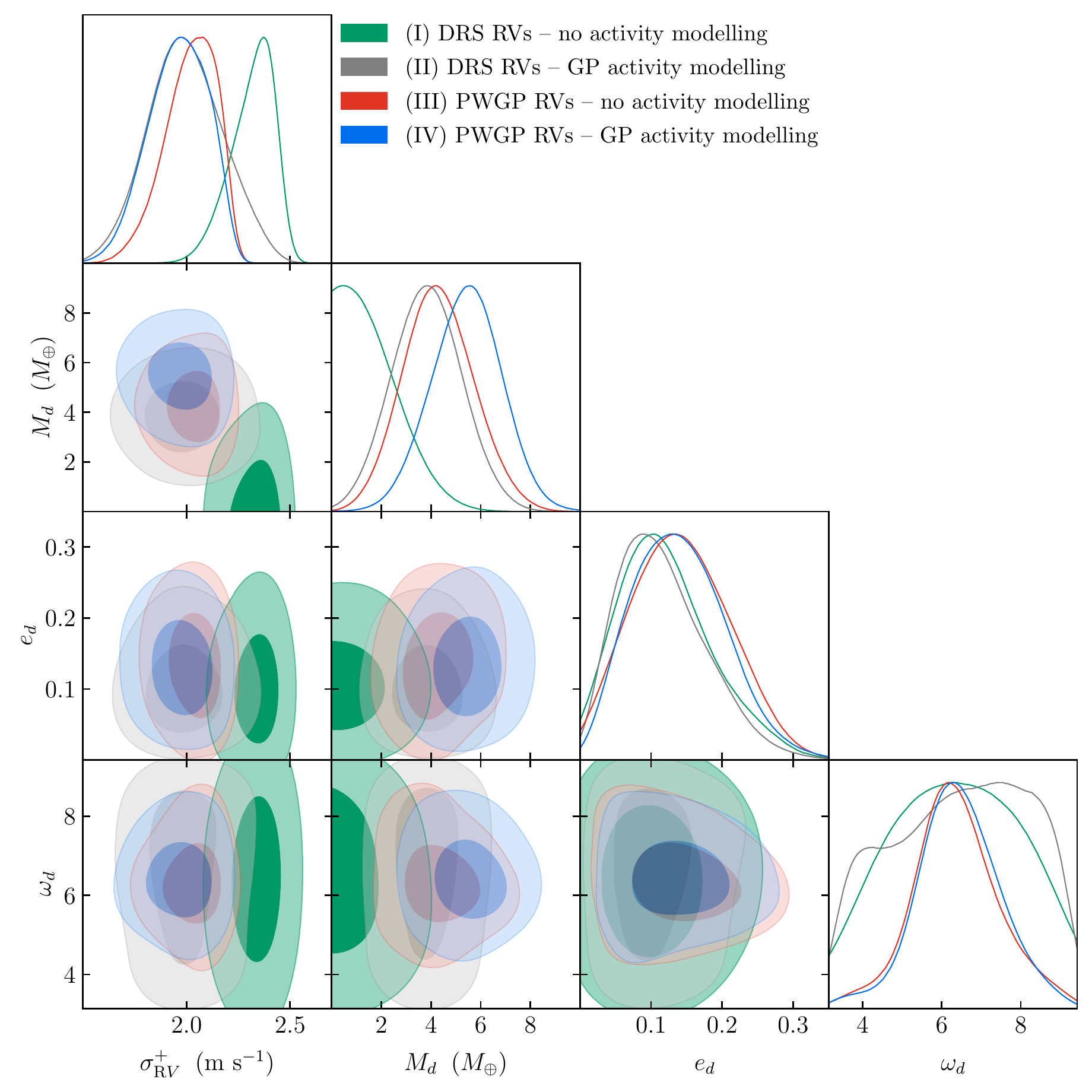}
    \caption{Corner plot showing marginalised 1D and 2D posterior probability densities for four key parameters shared between modelling approaches (I)--(IV), in each case where Kepler-37d was the only Keplerian component in the model. The dark and light filled regions correspond, respectively, to $1\sigma$ ($39.3\%$) and $2\sigma$ ($86.5\%$) joint credible regions. $\omega_d$ has been transformed to the domain $[\upi,3\upi]$ to suppress boundary effects, per the caption to Table~\ref{tab:param-posteriors}.}
    \label{fig:K37d_posterior}
\end{figure*}

\begin{figure*}

	\includegraphics[width=\textwidth]{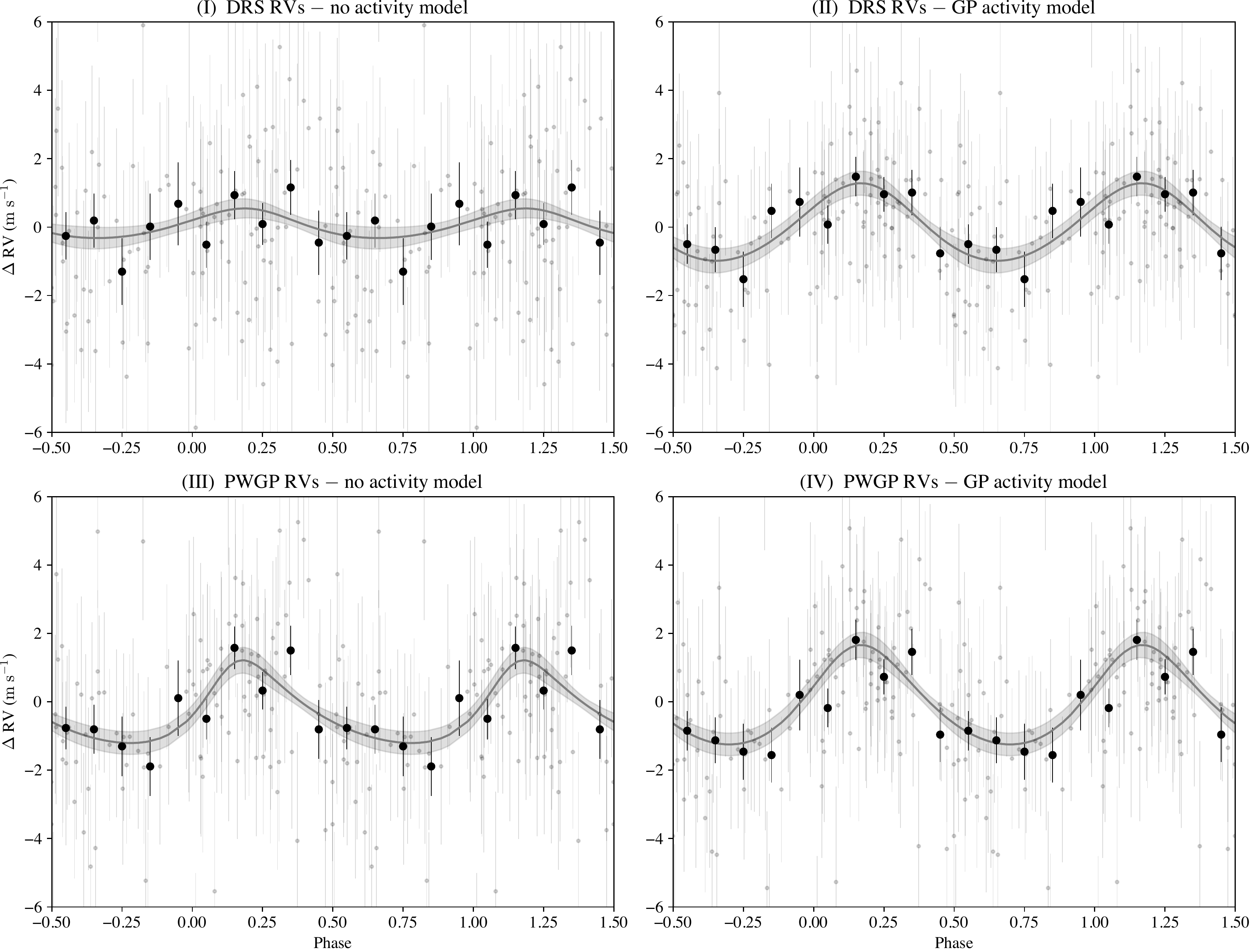}
    \caption{Kepler-37 RVs folded to the orbital period of Kepler-37d. The grey points show the DRS RVs (top panels) or PWGP RVs (bottom panels), either directly from the extraction pipelines (left panels) or after subtracting the best-fitting GP activity model (right panels); the associated error bars are the pipeline $1\sigma$ error bars. The solid grey lines indicate MAP models for Kepler-37d in approaches (I)--(IV), with the shaded grey regions indicating $\pm\sigma$ predictive uncertainty, as derived from the posterior uncertainty in the orbital parameters. Finally, the black points represent the RVs averaged into ten equally-spaced phase bins; here, the associated error bars are the standard errors on the mean RV of the points in each bin.}
    \label{fig:K37d_folded}
\end{figure*}

% Figure: GP posterior
\begin{figure*}
	\includegraphics[width=\textwidth]{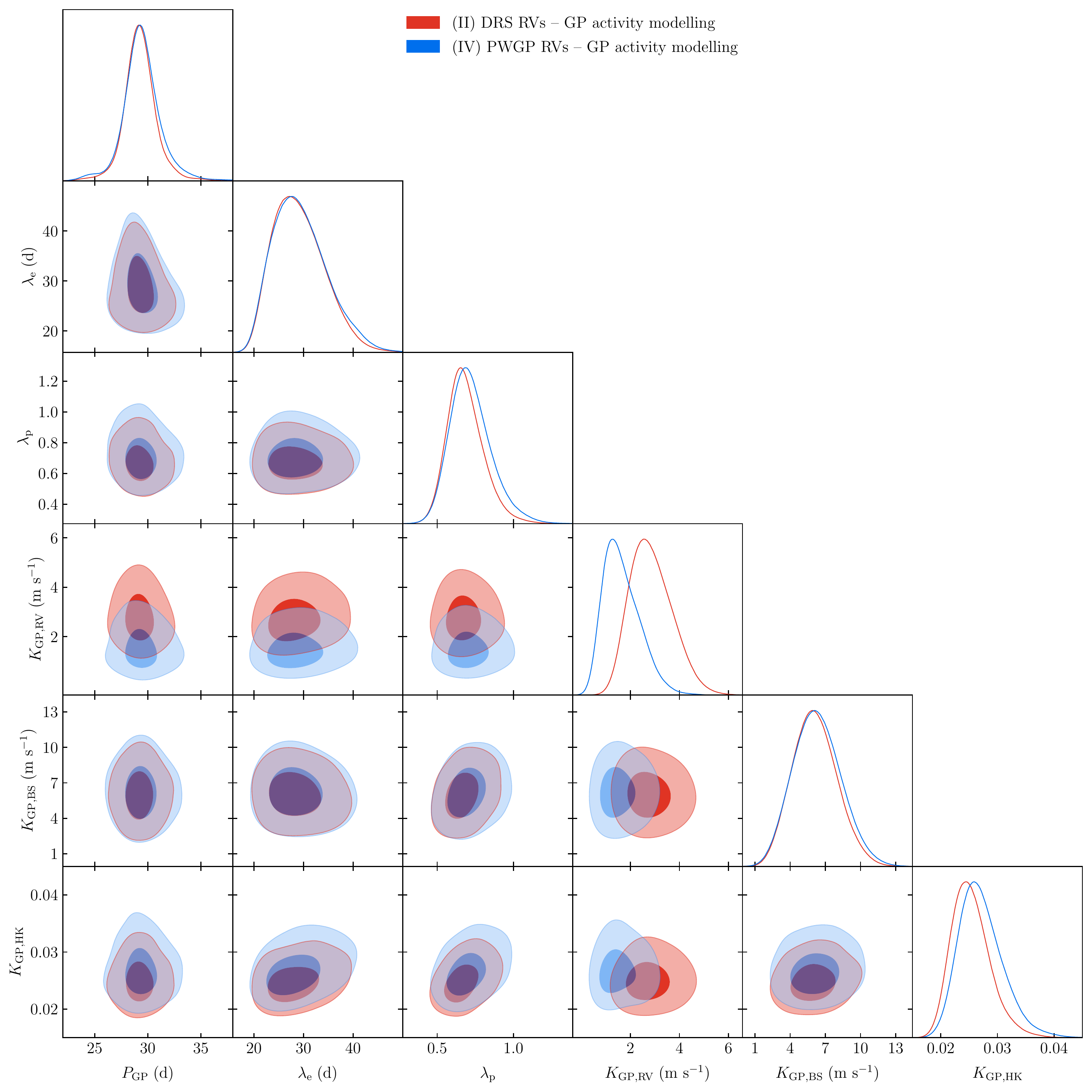}
    \caption{Corner plot as in Fig.~\ref{fig:K37d_posterior}, but now for GP-related (hyper-)parameters in the two GP-based modelling approaches, in each case where Kepler-37d was the only Keplerian component. To simplify interpretation and suppress sign degeneracy bimodalities inherent in the parameters $V_c$, $V_r$, $B_c$, $B_r$ and $L_c$, we plot the following (positive semi-definite) semi-amplitudes: $K_{\rm{GP,RV}}\equiv(V_r^2+V_c^2)^{1/2}$, $K_{\rm{GP,BS}}\equiv(B_r^2+B_c^2)^{1/2}$, $K_{\rm{GP,HK}}\equiv|L_c|.$}
    \label{fig:GP_posterior}
\end{figure*}

\begin{figure*}

	\includegraphics[width=\textwidth]{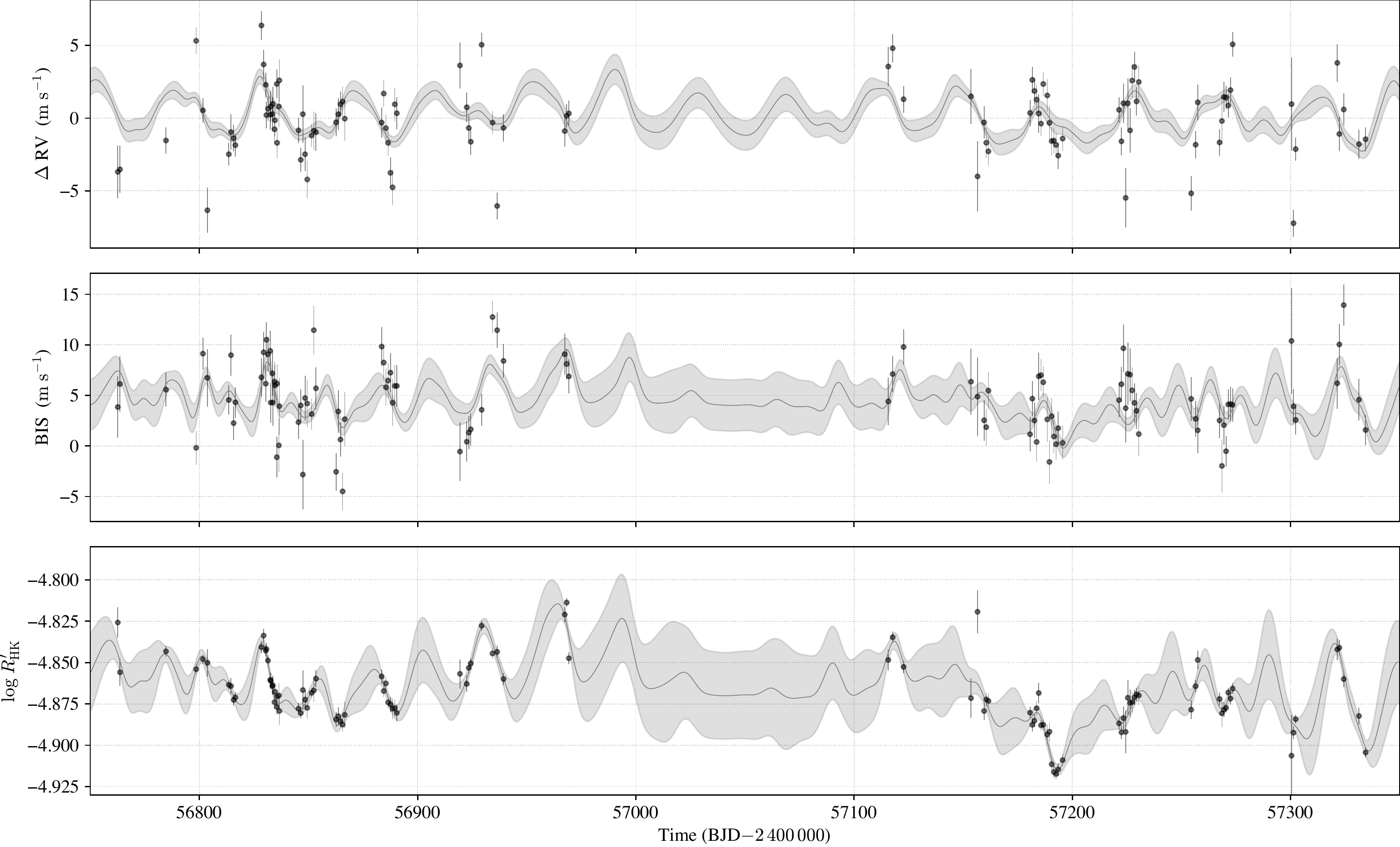}
    \caption{Joint fits to RV, BIS and \lrhk\ time series under approach (IV), for the model containing Kepler-37d but no other planets -- i.e., the model with the highest marginal likelihood. The dark grey points indicate observed values (with error bars); the solid grey lines indicate GP posterior predictive means (plus best-fitting Keplerian for the RVs), and the shaded regions denote $\pm\sigma$ posterior predictive uncertainties. Three observations from 2019 are not visible here.}
    \label{fig:GP_fit1}
    
\end{figure*}

\subsection{Parameter inference}\label{sec:param-posterior}

%We now shift our attention to the \emph{parameters} inferred under our favoured models. 
As the model with the highest evidence for both the DRS and PWGP RVs contained Kepler-37d as the sole Keplerian component, we focus here on the stellar and planetary parameters inferred under modelling approaches (I)--(IV), in each case where Kepler-37d was the only Keplerian term in the model.

Table~\ref{tab:param-posteriors} summarises the marginalised, 1D posteriors for all parameters in modelling approaches (I)--(IV). We highlight here a few of the most salient findings.

Crucially, the parameter posteriors for Kepler-37d show strong, reassuring consistency across all four modelling approaches -- with the exception of the RV semi-amplitude $K_d$, which is consistent within $1\sigma$ between approaches (II)--(IV) only, as Kepler-37d was not detected under approach (I). Far from the GP `eating up' a planetary signal (a concern often voiced in the RV community), our careful use of GP modelling across multiple time series allowed a subtle planetary signal to be recovered from correlated stellar nuisance signals. Indeed, the RV additive white noise parameter, $\sigma^+_{\rm RV}$, is consistent within $1\sigma$ between approaches (II)--(IV), but significantly larger in approach (I), where Kepler-37d's signal evidently remains `buried' behind stellar nuisance signals that were modelled as white noise. Given the consistency between posteriors from approaches (II)--(IV), we avoid the temptation to over-interpret any supposed differences between the results, e.g.\ the marginally smaller error bars on $K_d$ in approach (III) vs.\ (IV). 

Fig.~\ref{fig:K37d_posterior} shows 1D and 2D marginalised posteriors for the aforesaid parameters, albeit with semi-amplitude $K_d$ converted into a more physically interesting mass $M_d$, across approaches (I)--(IV). Meanwhile Fig.~\ref{fig:K37d_folded} shows \mystar\ RVs folded to the orbital period of \mystar d, after subtracting applicable stellar activity models, again across approaches (I)--(IV).

Turning to the activity modelling, Table~\ref{tab:param-posteriors} shows that, unsurprisingly, $\sigma^+_{\rm BS}$ and $\sigma^+_{\rm HK}$ end up significantly larger in approaches (I) and (III) than in (II) and (IV), since in the latter two cases, the GP allows variability in the BIS and \lrhk\ time series to be modelled as correlated signals (which they certainly are), rather than white noise.

The quasi-periodic GP hyper-parameters are consistent between approaches (II) and (IV); they suggest a stellar rotation period of $P_{\rm GP}=29\pm1$~d, and an active region evolution time scale of order one rotation period, i.e.\ $\lambda_{\rm e}=29^{+1}_{-4}$~d. The hyper-parameter $\lambda_{\rm p}\sim0.7$ suggests the stellar RV signals modelled were of moderate harmonic complexity, having roughly twice as many turning points per period as a sine wave \citep{vinesh-thesis}. The relatively tight constraint on the stellar rotation period under the GP model contrasts strongly with the weak constraints from GLS analyses of activity indicators (see e.g.\ Fig.~\ref{fig:K37_Lomb_data}); these findings are unsurprising, though, given that the activity signal is both non-sinusoidal and fairly rapidly evolving, and therefore inherently unsuitable for characterisation via a GLS periodogram.

Importantly, we found that the GP's RV covariance amplitudes ($V_r$, $V_c$) ended up about a factor of two smaller in approach (IV) than (II), though the amplitudes for the other time series ($B_r$, $B_c$, $L_c$) were consistent across the two approaches. This is compelling evidence that our efforts to mitigate RV activity contamination via the PWGP approach were successful, albeit imperfect -- the latter also evidenced by the improvements seen when moving from approach (III) to (IV). The fact that $|V_r|>|V_c|$ and $|B_r|>|B_c|$ suggests the stellar signals we modelled arose primarily from rotating active regions, rather than their associated suppression of convective blueshift \citep{aigrain2012,Rajpaul2015}.

Fig.~\ref{fig:GP_posterior} shows 1D and 2D marginalised posteriors for the aforementioned GP (hyper-)parameters, across approaches (I)--(IV), while our best-fitting model from approach (IV), including the simultaneous GP fit to the RV, BIS and \lrhk\ time series, is shown in Fig.~\ref{fig:GP_fit1}. 

\hl{While the \lrhk~series is clearly dominated by oscillations with a $\sim30$~d (quasi-)period, it also shows some evidence for a long-term trend, with the median values in the 2014 season being higher than those in the 2015 season. Though not visible in Fig}.~\ref{fig:GP_posterior}, \hl{this apparent trend continued in 2019, when $\lrhk\le-4.93$ in all three spectra taken that year, suggesting the star was moving to a lower-activity phase over the years we observed it. A similar trend was evident in the FWHM series. These trends correspond to some of the lower-frequency structures in the \lrhk~and FWHM periodograms in Fig.}~\ref{fig:K37_Lomb_data}\hl{, and may be evidence of a longer-term stellar magnetic cycle.}

\hl{It is also evident, in Fig}.~\ref{fig:GP_fit1}, \hl{that the residual scatter from the GP fit to the \lrhk~series is smaller than in the RV and BIS cases. This was to be expected \emph{a priori}, given that the \lrhk~error bars are about half the size of the RV or BIS error bars, when considered as a fraction of the rms variation in the respective time series. This difference is further amplified when factoring in the proportionately larger RV and BIS white noise jitter terms (see Table}~\ref{tab:param-posteriors}).

\subsection{Dynamical stability analysis}\label{sec:stability}

We used the code \texttt{Mercury-T} \citep{mercury-T} to run full dynamical simulations of the Kepler-37 system, including general relativistic precession and precession due to stellar rotational flattening. Using the parameter values from approach (IV) in Table~\ref{tab:param-posteriors}, we found that the system was unstable over $10^6$ years, mostly through the eccentricity excitation and ejection of Kepler-37b; this did not depend on the inclusion of candidate planet e. However, when we lowered Kepler-37d's mass and eccentricity to the lower end of our $1\sigma$ posterior credible interval ($M_d=4$~$M_\oplus$, $e_d=0.075$) and assumed Kepler-37b was iron-rich ($K_b\sim0.6$~\cmps, $M_b\sim0.03$~$M_\oplus$), the system generally remained stable over $10^6$ years, again regardless of whether Kepler-37e was included.

We are obliged to conclude that Kepler-37d's RV semi-amplitude and/or eccentricity are on the lower end of our inferred values, or that Kepler-37b's RV semi-amplitude is higher than our MAP value ($K_b\sim0.3$~\cmps). However, our RVs essentially do not constrain the orbital parameters of Kepler-37b -- their posteriors being overwhelmingly shaped by our broad priors -- yet \emph{do} place strong constraints on the orbit of Kepler-37d. Under all modelling approaches, we ended up with a moderate eccentricity ($e_d\sim0.12$ or $e_d\sim0.14$) inconsistent with zero at a $2$--$3\sigma$ level. As such, we are inclined to favour Kepler-37b being iron-rich (\tocheck{perhaps having ejected much of a former silicate mantle in a collision}), rather than Kepler-37d's orbit having negligible eccentricity. In any case, whether $e_d$ is zero or moderately large has negligible impact on the mass $M_d$ inferred from $K_d$. 

There may \emph{also} be an orbital configuration that could stabilise the whole system, even with Kepler-37d being relatively massive and on a significantly eccentric orbit.  However, exploring a much wider range of parameter space is beyond the scope of this paper (cf.\ comments about computational burdens in Section~\ref{sec:burden}). In the  future, joint GP activity modelling and dynamical simulations would likely help refine the orbital parameters for Kepler-37d, if not the other transiting planets.

\subsection{Kepler-37e: re-examining Kepler-37d's TTVs}

%The presence of a planet with $\sim51$~d period was claimed by \citet{hadden+14}, based on a TTV analysis of the first $12$ quarters of the Kepler mission. 
%As we noted in Section~\ref{sec:k37e?}, there are reasons to doubt the reality (or at least planet-like nature) of the signal ascribed by \citet{hadden+14} to what is currently referred to as Kepler-37e. 

We carried out a separate dynamical analysis and re-examination of the Kepler-37 TTVs, hoping to shed light on the nature of the $\sim51$~d period signal currently ascribed to Kepler-37e.

% \subsubsection{TTV analysis}
%\subsubsection{Probing TTVs}

\begin{figure}
\centering
	\includegraphics[width=\hsize]{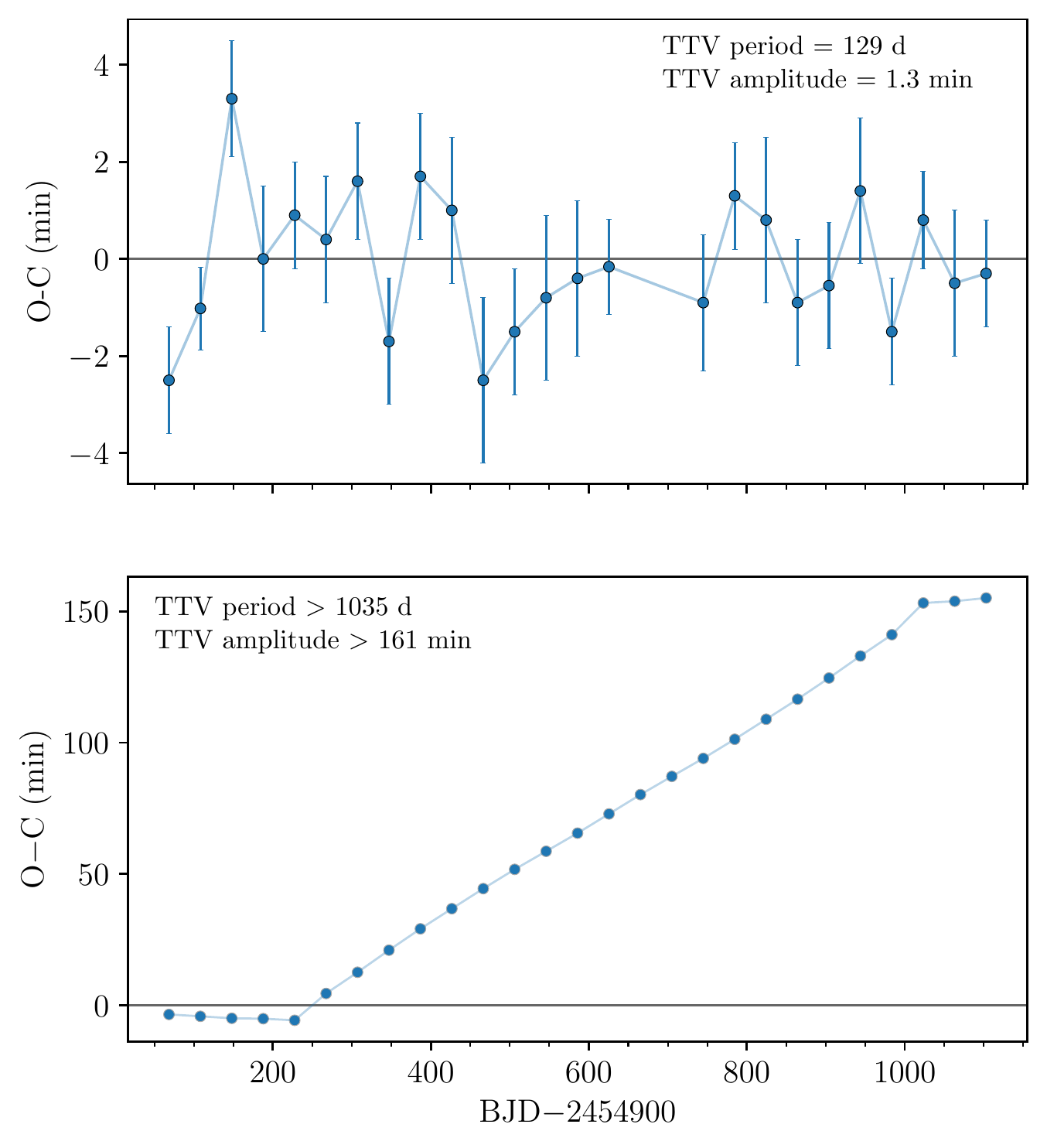}
    \caption{\emph{Top}: measured TTV signal of Kepler-37d according to the transit times reported in \citet{mazeh13}. The TTV period and amplitude are computed via the GLS periodogram \citep{zechmeister2009}. \emph{Bottom}: predicted TTV signal of Kepler-37d according to our numerical simulation, assuming the presence of a planet with $\sim51$~d period. The simulation suggests a TTV period longer than the observed baseline, and amplitude $\gtrsim 160$ min.
}
    \label{fig:ttv_k37d}
\end{figure}

The periods of Kepler-37d and e are close to a 4:3 commensurability, hinting at a first-order mean motion resonance, which may suggest the presence of a detectable TTV signal induced on planet d due to strong dynamical interactions between the two planets. Consequently, by studying the TTV signal of Kepler-37d, we can infer some information on the hypothetical perturbing planet, i.e.\ Kepler-37e. According to \citet{mazeh13}, whose transit times were used in the analysis by \citet{hadden+14}, Kepler-37d has a shallow TTV signal with an amplitude of order one minute, as reported in the observed minus calculated diagram (O$-$C, \citealt{agol2018}) in Fig.~\ref{fig:ttv_k37d}. 

We numerically integrated the orbits of Kepler-37b, c, d and e using the $N$-body integrator \texttt{ias15} within the {\tt rebound} package \citep{rein2012}, assuming as reference time the transit mid-point time ($T_\mathrm{0,d}$) of Kepler-37d. As initial configuration, we assumed the planetary parameters in Table~\ref{tab:planet-properties}, except for the mass and eccentricity of Kepler-37d, for which we used our inferred values (Table~\ref{tab:planet-derived}). 
For planet e, since \citet{hadden+14} only report the planetary period, we used the mid-transit time as originally reported by \citet{batalha2013}, $T_0 = 2455028.727 \pm 0.0096$ $\rm{BJD}_{\rm TDB}$, and we derived the radius from the planet-to-star radius ratio given by \citet{batalha2013}: $R_\mathrm{e}/R_\star = 0.0054 \pm 0.0002$, $R_\mathrm{e} = 0.43 \pm 0.03$~$R_\oplus$. We estimated the masses of planets b, c, and e using the \citet{Wolfgang_2016}\footnote{\url{https://github.com/dawolfgang/MRrelation}.} probabilistic mass-radius relation: $M_\mathrm{b} = 0.01$~$M_\oplus$, $M_\mathrm{c} = 0.6$~$M_\oplus$, $M_\mathrm{e} = 0.06$~$M_\oplus$. 
During the integration, we computed the synthetic transit times (O) of each planet following the procedure described in \cite{borsato2019}, and compared the inferred transit times with the linear ephemeris (C) of \citet{mazeh13}, obtaining the synthetic O$-$C diagram to identify a possible TTV signal.

Despite the low expected mass of the putative planet e, the predicted TTV signal induced on Kepler-37d has a high amplitude, due to the suggested resonant configuration of the planets and to the eccentricity of planet d. Fig.~\ref{fig:ttv_k37d} shows that neither the amplitude nor the period of such TTV signal corresponds to the observed TTVs.

It is worth noting that the TTV prediction is highly dependent on the planetary parameters, and in particular, even small changes in the eccentricity of Kepler-37d imply a variation of TTV amplitude from $0$ to more than $160$ minutes. However, our simulations predict that in order to obtain a TTV amplitude of order of the observed one, the eccentricity of Kepler-37d would need to be negligible ($e_\mathrm{d} \leq 0.01$), whereas our RV modelling points towards a non-trivial eccentricity. 

We also computed the same forward modelling assuming a 3-planet system, i.e. without the presence of Kepler-37e. In this case, the amplitude of the predicted TTV signal of Kepler-37d was of order $0.2$ minutes, that is, lower than the average error on the transit times of \citet{mazeh13}, indicating that the two inner planets do not significantly perturb the orbit of planet d.

In summary, our TTV analysis disfavours the presence of the putative planet Kepler-37e. Considering the dubious way in which Kepler-37e became a `confirmed' planet in the first place (see Section~\ref{sec:k37e?}), the fact that our RV analysis did not lead to the detection of a signal with $\sim51$~d period, the prominence of the $\sim51$~d periodicity in two activity indicators, and the additional doubts introduced by our TTV forward modelling, we suggest that Kepler-37e be stripped of its status of a `confirmed' planet.

We do not exclude the possibility of a non-transiting planet inducing the small observed TTV signal of Kepler-37d, which cannot be totally accounted for by a 3-planet system, but the properties of such a non-transiting planet would not seem to correspond to those of Kepler-37e as reported in the literature.

\subsection{Final characterisation of transiting planets}\label{sec:results-K37d}

% \rho_\oplus=5.51$~g~cm$^{-3}$

\begin{figure}
\centering
	\includegraphics[width=\linewidth]{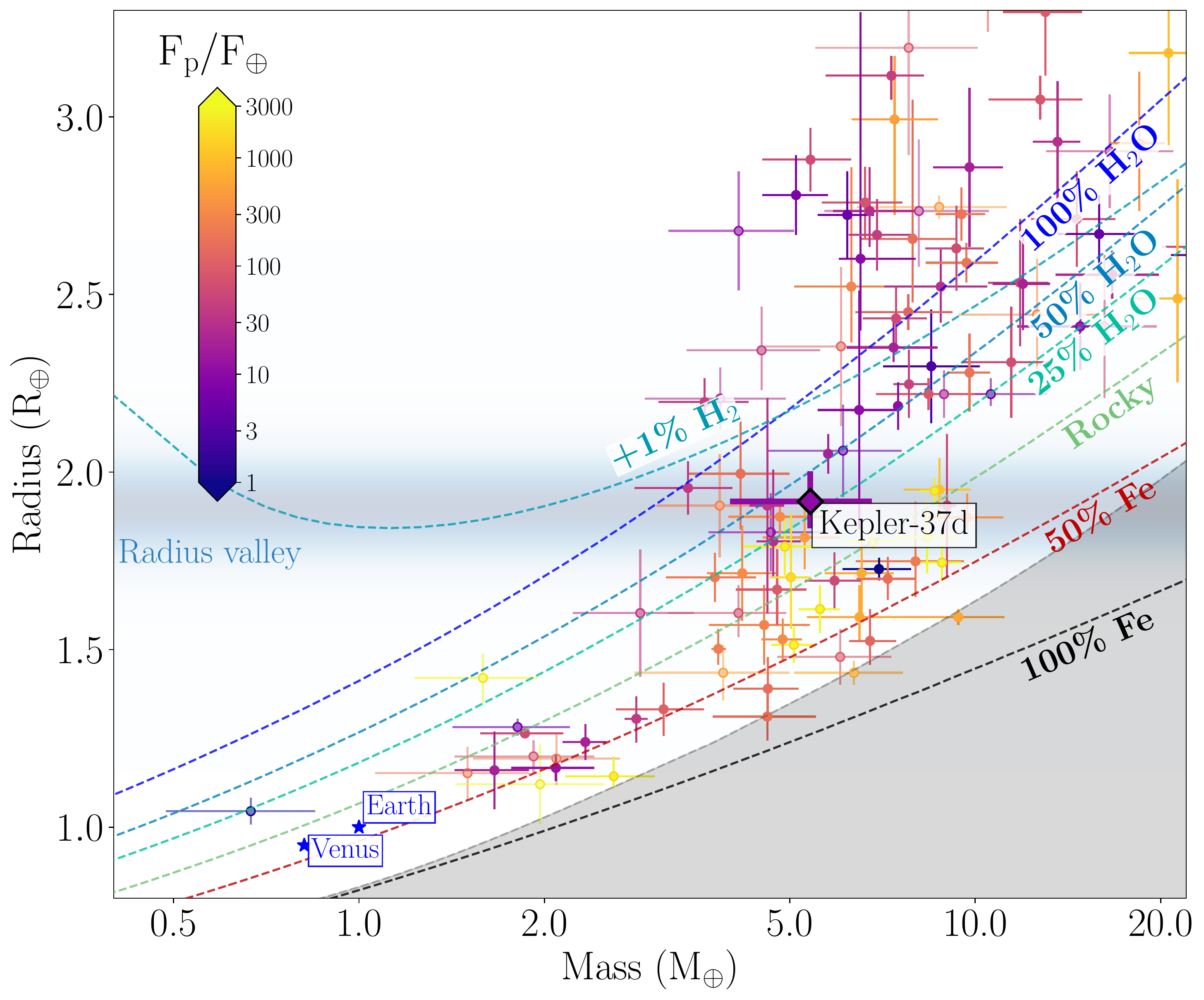}
    \caption{Mass-radius diagram for exoplanets with masses and radii measured with better than $30$\% precision, colour-coded according to their incidental flux (Earth units). Kepler-37d is indicated with a large diamond. The dashed coloured lines represent theoretical mass-radius curves for various chemical compositions according to \citet{zeng19}. The shaded grey region denotes the maximum Fe content predicted by collisional stripping \citep{marcus2010}. Data taken from the The Extrasolar Planets Encyclopaedia (\url{http://exoplanet.eu/catalog/}) on 2021 March 26.
}
    \label{fig:MR_diagram}
\end{figure}

In Table~\ref{tab:planet-derived} we summarise our final fitted and derived parameters for Kepler-37's transiting planets -- for Kepler-37d, drawn from our highest-evidence model from approach (IV), and for planets b and c, drawn from the best-fitting model including these Keplerians under approach (IV). We assume Kepler-37e is not present, and disregard any ambiguities raised by our dynamical modelling (Section~\ref{sec:stability}).

\begin{table}
	\centering
	\caption{Selected fitted and derived parameters for the three planets known to transit \mystar. Orbital periods, radii, and orbital inclinations in Table~\ref{tab:planet-properties} were used to derive planetary masses and densities. The parameter posteriors for Kepler-37b were essentially the same as our priors.}
	\label{tab:planet-derived}
\begin{tabular}{lcc}
\hline
Parameter & Posterior & Notes \\  
\hline
$K_b$ (\mps) & $<0.007$ & $95\%$ upper limit \\
$M_b$ ($M_\oplus$) & $<0.02$ & $95\%$ upper limit \\
$\rho_b$ (g~cm$^{-3}$) & $<6$ & $95\%$ upper limit \\
\hline
$K_c$ (\mps) & $<0.1$ & $95\%$ upper limit  \\
$M_c$ ($M_\oplus$) & $<0.6$ & $95\%$ upper limit \\
$\rho_c$ (g~cm$^{-3}$) & $<5$ & $95\%$ upper limit \\
\hline
$K_d$ (\mps) & $1.22\pm0.31$ & $3.9\sigma$ detection\\
$e_d$ &  $0.142^{+0.026}_{-0.044}$ & \bl \\
$a_d$ (AU) & $0.2109\pm 0.0029$ & Orbital semi-major axis \\
$M_d$ ($M_\oplus$) & $5.4\pm1.4$ & $3.9\sigma$ detection \\
$\rho_d$ (g~cm$^{-3}$) & $4.29^{+0.52}_{-0.74}$ & \bl \\
\hline
\end{tabular}
\end{table}

In Fig.~\ref{fig:MR_diagram} we show the position of Kepler-37d in the mass-radius diagram of exoplanets with masses and radii both measured with better than $30$\% precision. 
According to our derived density ($\rho_d=4.29^{+0.52}_{-0.74}$~g~cm$^{-3}$), the interior composition of Kepler-37d differs from that of an Earth-like planet (around $30\%$ Fe and $70\%$ silicates) at a $> 2\sigma$-level in both radius and mass. Instead, its density is consistent with a $\sim 25$\% H$_2$O composition, making it a compatible with a water-world scenario, where \lq water worlds\rq\ are  planets with massive water envelopes, in the form of high pressure H$_2$O ice, comprising $>5\%$ of the total mass. However, its density could also be explained via a gaseous envelope surrounding a rocky core.  Following \citet{Lopez_2014}, assuming a rocky Earth-like core, a Solar composition H-He envelope, and using our derived planetary and stellar parameters, we estimate that an H-He envelope comprising $\sim 0.4$\% of the planet mass could also explain the observed properties of Kepler-37d. 

On the other hand, if Kepler-37d actually has a mass closer to $M_d\sim4$~$M_\oplus$ (as our dynamical modelling tentatively suggests could be the case), its consequent density $\rho_d\sim3.2$~g~cm$^{-3}$ would be more consistent with a $\sim50\%$ water composition.

We note that Kepler-37d lies within the so-called small planet radius gap or `Fulton gap' \citet{fulton17}. According to recent XUV irradiation modelling by \citet{galian20}, which reproduces the bimodal exoplanet radius distribution observed by Fulton, roughly equal proportions of planets with Kepler-37d's $1.9$~$R_\oplus$ radius are expected to have atmospheres vs no atmospheres.

%\subsubsection{Conclusion}

\section{Conclusions}\label{sec:discuss}

With an RV semi-amplitude of $1.22\pm0.31$~\mps, Kepler-37d is one of only a handful of transiting planets securely detected below a $2$~\mps\ RV threshold. Its RV semi-amplitude is only slightly larger than (though formally consistent with) TOI-178b's  $1.05^{+0.25}_{-0.30}$~\mps, the latter RV signal detected with ESPRESSO, and currently the smallest of any transiting planet detected to date. By extension, Kepler-37d has (to our knowledge) the smallest RV semi-amplitude of any transiting planet characterised with HARPS-N alone, or indeed with any single second-generation spectrograph. Based on our RV modelling alone, our best estimate of Kepler-37d's mass is $5.4\pm1.4$~$M_\oplus$, although full dynamical simulations suggest its mass might be on the lower end of that $1\sigma$ credible interval.

As expected, we did not detect either Kepler-37b or c, whose RV semi-amplitudes lie far below the threshold of current detectability. Our RV modelling and re-examination of Kepler-37's TTVs suggested that the putative planet Kepler-37e should probably be stripped of its `confirmed planet' status.

Stepping back from the specifics of the \mystar\ system, our analysis served as a case study in stellar activity modelling and mitigation. We showed how (i) careful modelling of stellar activity in HARPS-N pipeline RVs and activity indicators, or (ii) mitigating stellar activity \emph{before} RV extraction, in either case using state-of-the-art tools, enabled a RV detection of a planet that had eluded secure RV characterisation for many years. The PWGP RV extraction technique we leveraged in the latter approach appears to have some advantages over the DRS; a systematic comparison of the DRS with many different extraction techniques -- in the vein of the EXPRES Stellar-Signals Project \citep{ESSP2020}, ideally including spectra with injected planetary signals -- could be useful to illuminate their relative merits.

Importantly, we demonstrated that the aforesaid approaches of activity mitigation and modelling not only yielded consistent results, but can be complementary: when combined, they led to a decisive detection and more precise characterisation of Kepler-37d. Such a two-pronged approach might be crucial for enabling the detection of Earth-twin exoplanets with next-generation surveys such as the Terra Hunting Experiment \citep{hall2018}.  

\section*{Acknowledgements}
\hl{We are grateful to the anonymous reviewer whose insights helped improve this work.} We also wish to thank Howard Isaacson and Courtney Dressing for their assistance interpreting the published HIRES RVs, and for providing newer RVs extracted using the latest HIRES pipeline. 

VMR acknowledges the Royal Astronomical Society and Emmanuel College for financial support, and the Cambridge Service for Data Driven Discovery (CSD3) -- operated by the University of Cambridge Research Computing Service (\url{www.csd3.cam.ac.uk}), provided by Dell EMC and Intel using Tier-2 funding from the Engineering and Physical Sciences Research Council (capital grant EP/P020259/1), and DiRAC funding from the Science and Technology Facilities Council (\url{www.dirac.ac.uk}) -- for providing computing resources. GL and LBo acknowledge funding from the Italian Space Agency (ASI) via `Accordo ASI-INAF n.\ 2013-016-R.0 del 9 luglio 2013 e integrazione del 9 luglio 2015 CHEOPS Fasi A/B/C'; GL also acknowledges support from CARIPARO Foundation, via agreement CARIPARO-Universit\`a degli Studi di Padova (Pratica n.\ 2018/0098). SA acknowledges funding from the UK Science and Technology Facilities (STFC) research council via consolidated grant ST/S000488/1, and from the European Research Council (ERC) via grant agreement n.\ 865624. AMo acknowledges support from the senior Kavli Institute Fellowships. FP %e %and CLo 
acknowledges the Swiss National Science Foundation (SNSF) for supporting research with HARPS-N through SNSF grants n.\ 140649, 152721, 166227 and 184618. 

This work is based on observations made with the Italian Telescopio Nazionale Galileo (TNG) operated on the island of La Palma by the Fundaci\'on Galileo Galilei of the INAF (Istituto Nazionale di Astrofisica) at the Spanish Observatorio del Roque de los Muchachos of the Instituto de Astrofisica de Canarias. The HARPS-N project was funded by the ESA-PRODEX Program of the Swiss Space Office (SSO), the Harvard University Origin of Life Initiative (HUOLI), the Scottish Universities Physics Alliance (SUPA), the University of Geneva, the Smithsonian Astrophysical Observatory (SAO), the Italian National Astrophysical Institute (INAF), University of St. Andrews, Queen's University Belfast, and University of Edinburgh.
%%%%%%%%%%%%%%%%%%%%%%%%%%%%%%%%%%%%%%%%%%%%%%%%%%
\section*{Data Availability}

%The inclusion of a Data Availability Statement is a requirement for articles published in MNRAS. Data Availability Statements provide a standardised format for readers to understand the availability of data underlying the research results described in the article. The statement may refer to original data generated in the course of the study or to third-party data analysed in the article.
All RVs and activity indicators used in our analyses will be available via VizieR at CDS.


\bibliographystyle{mnras}
\bibliography{biblio}

%%%%%%%%%%%%%%%%%%%%%%%%%%%%%%%%%%%%%%%%%%%%%%%%%%

%%%%%%%%%%%%%%%%% APPENDICES %%%%%%%%%%%%%%%%%%%%%

\appendix

\section{Keck-HIRES radial velocities}\label{sec:HIRES}

There exist 33 publicly-available Keck-HIRES RVs for Kepler-37 \citep{barclay+13}, extracted from spectra taken between 2010 and 2012, with 18 of them taken in 2011 August. Their RMS is $3.05$~\mps, and mean error bar $1.44$~\mps. Using these HIRES RVs, \citet{marcy2014} derived the following masses for Kepler-37's transiting planets ($95\%$ upper limits in parenthesis): $M_b=2.8\pm3.7$ ($10.0$)~$M_\oplus$, $M_c=3.4\pm4.0$ ($12.0$)~$M_\oplus$, $M_d=1.9\pm9.1$ ($12.2$)~$M_\oplus$. At face value, these HIRES masses are all consistent with zero.

We ran our models from approaches (I) and (II) using the HIRES RVs alone, as well as HARPS-N plus HIRES RVs. However, we ultimately omitted these thornier analyses from the body of our paper for several reasons. First, the PWGP RV extraction method is not designed for iodine-cell spectra; even if we had the original HIRES spectra at our disposal, we could not have investigated activity mitigation (a primary focus of this study) with those spectra. Second, the HIRES RVs were accompanied by Mount-Wilson $S$-index measurements only, but no CCF-derived indicators such as the BIS. Third, joint modelling of RVs and activity indicators from both instruments required significantly more free parameters, to accommodate instrumental offsets, independent additive white noise terms, etc. The HIRES $S$-index series also evinced quadratic-like variability over its three-year span, whereas the HARPS-N activity indicators did not.

The upshot from approaches (I) and (II) was that we did not detect Kepler-37d in the HIRES RVs, and made only a doubtful detection when combining them with the HARPS-N RVs under approach (II). In the latter case, we inferred $K_d=0.6\pm0.28$~\mps, albeit with $\mathcal{R}_{d0}\gtrsim1$. Re-running all our models with HIRES RVs extracted with the newest pipeline available in 2020 July did not change our conclusions. This seems to parallel the confounding situation in which HIRES RVs, HARPS-N RVs, and their combination each implies discrepant masses for Kepler-10c \citep[e.g.][]{Rajpaul2017}.

We may speculate about why the HIRES RVs hinder our characterisation of Kepler-37d (which, unlike Kepler-10c, our HARPS-N RVs indicate to have unexceptional properties). The HIRES spectra were taken years before the first HARPS-N spectra, when \mystar\ exhibited significantly more activity-related variability (based on $S$-index measurements); stronger stellar signals in the HIRES RVs may exert disproportionate influence on model fits. The HIRES and HARPS-N spectra cover slightly different wavelength ranges ($360$--$800$~nm versus $383$--$690$~nm), so stellar signals measured by each instrument could have different properties -- something our modelling did not account for -- regardless of when the star is observed. 

%Or perhaps the many extra parameters required to model observations from the different instruments introduced enough flexibility into the global model that Kepler-37d's subtle signal could no longer be safely identified as Keplerian.

\hl{Instrumental systematics peculiar to HIRES are another possibility. We note that 8 of the 33 HIRES observations were made using the B5 decker/slit combination, with the remainder made using the C2 decker/slit combination. While the modes share a $R=48000$ resolution and $0.861\arcsec$ slit width, their slit lengths differ, the former being $3.5\arcsec$ vs $14\arcsec$ for the latter. Ongoing investigations suggest HIRES RVs taken with the B5 aperture may be associated with significantly higher RV scatters, at least in the case of Kepler-10. Thus, a relatively small number of outlying HIRES RVs, with under-estimated error bars, might be dominating model fits.}

Whatever the case, considering that many hundreds of RVs and ongoing analyses have still not reconciled HIRES and HARPS-N RVs for Kepler-10, attempting to do so with Kepler-37 was well beyond the scope of this study. Fortunately, our detection of Kepler-37d was straightforward and secure using HARPS-N observations alone.

% 31 observations with B5 decker; balance of 110 with C2 - see file KOI245_HIRES_2021_07_06
%%%%%%%%%%%%%%%%%%%%%%%%%%%%%%%%%%%%%%%%%%%%%%%%%%

% Don't change these lines
\bsp	% typesetting comment
\label{lastpage}
\end{document}